\documentclass[twocolumn,english,aps]{revtex4-1}
\usepackage[T1]{fontenc}
\usepackage[latin9]{inputenc}
\usepackage{babel}
\usepackage{amsmath}
\usepackage{amssymb}
\usepackage{graphicx}
\usepackage[unicode=true]
 {hyperref}
\usepackage{breakurl}
\makeatletter
\usepackage{titlesec}
\usepackage[below]{placeins}
\usepackage{color}
\titlespacing*{\subsubsection}
{0pt}{1.5\baselineskip}{1\baselineskip}
\titlespacing*{\section}
{0pt}{2\baselineskip}{1\baselineskip}
\makeatother

\begin{document}

\title{Quantum phases of a three-level matter-radiation interaction
model using $SU(3)$ coherent states with different cooperation numbers}

\author{L.F. Quezada}

\email{luis.fernando@correo.nucleares.unam.mx}

\affiliation{Instituto de Ciencias Nucleares, Universidad Nacional Aut\'onoma de
M\'exico, Apartado Postal 70-543, 04510 Ciudad de M\'exico, M\'exico.}

\author{E. Nahmad-Achar}

\email{nahmad@nucleares.unam.mx}

\affiliation{Instituto de Ciencias Nucleares, Universidad Nacional Aut\'onoma de
M\'exico, Apartado Postal 70-543, 04510 Ciudad de M\'exico, M\'exico.}

\begin{abstract}
We use coherent states as trial states for a variational approach
to study a system of a finite number of three-level atoms interacting
in a dipolar approximation with a one-mode electromagnetic field.
The atoms are treated as semi-distinguishable using different cooperation
numbers and representations of SU(3). We focus our analysis on the
quantum phases of the system as well as the behavior of the most relevant
observables near the phase transitions. The results are computed for
all three possible configurations ($\Xi$, $\Lambda$ and $V$) of
the three-level atoms.
\end{abstract}
\maketitle

\section*{Introduction}

The study of the coherence in the radiation from a system of two-level identical atoms interacting with a one-mode quantized electromagnetic field was first described by Dicke \cite{key-1}. Dicke's model can be generalized to study systems of multiple-level atoms, allowing meaningful interactions with more modes of the electromagnetic field. In particular, systems of three and four-level atoms have been extensively studied \cite{key-9, key-10, key-11, key-12, key-13, key-14, key-15, key-16, key-17, key-18, key-19} as they have been proved useful in the development of certain types of quantum memories \cite{key-20, key-21, key-22, key-23}.

Two of the major aspects of these matter-radiation interaction models are the existence of quantum phase transitions (QPT's) and the distinguishability of the atoms. QPT's are informally seen as sudden, drastic changes in the physical properties of the ground state of a quantum system at zero temperature due to the variation of some parameter involved in the modeling Hamiltonian, while the distinguishability of the atoms is a characteristic that depends on the space we choose for the Hamiltonian to act on.

In this work, in order to study its QPT's, we use a variational approach to estimate the ground state of a system of a finite number of three-level atoms interacting in a dipolar approximation with a one-mode electromagnetic field. Most works on the subject treat the atoms as completely indistinguishable; this, however, may not correctly describe some of the experimental realizations of the models. To gain distinguishability we add information of the atomic field to the states we use to describe it, this information is the \textit{cooperation number}, a quantity closely related to the group's representation of the atomic field.

\subsubsection*{Modeling Hamiltonian}

The Hamiltonian describing the interaction, in a dipolar approximation, between $N$ three-level identical atoms (same energy levels) and one-mode of an electromagnetic field in an ideal cavity, has the expression ($\hbar=1$) \cite{key-16}

\begin{multline}
H=\overline{\omega}_{1}e_{11}+\overline{\omega}_{2}e_{22}+\overline{\omega}_{3}e_{33}+\Omega a^{\dagger}a\\-\frac{1}{\sqrt{N}}\sum_{i<j}^{3}\mu_{ij}\left(e_{ij}+e_{ij}^{\dagger}\right)\left(a+a^{\dagger}\right).
\label{eq:1}
\end{multline}

Here, $\overline{\omega}_{1}$, $\overline{\omega}_{2}$ and $\overline{\omega}_{3}$ are the three energy levels of the atoms, with $\overline{\omega}_{1} \leq \overline{\omega}_{2} \leq \overline{\omega}_{3}$, $\Omega$ is the frequency of the field's mode, $\mu_{ij}$ are the dipolar coupling parameters between the radiation and the pair of atomic levels $i$ and \mbox{$j$, $a$ and $a^{\dagger}$} are the annihilation and creation operators of the harmonic oscillator, and $e_{ij}$ are the collective atomic matrices (annihilation operators for the atomic field), i.e., summations (with as many summands as atoms in the system) of the single-entry matrices $\left(\overline{e}_{ij}\right)_{mn}=\delta_{im}\delta_{jn}$. Choosing the zero of the energy to be at $\frac{1}{3}\left(\overline{\omega}_{1}+\overline{\omega}_{2}+\overline{\omega}_{3}\right)$ we can rewrite this hamiltonian (\ref{eq:1}) in the more useful form

\begin{multline}
H=\omega_{1}J_{z}^{\left(1\right)}+\omega_{2}J_{z}^{\left(2\right)}+\Omega a^{\dagger}a\\-\frac{1}{\sqrt{N}}\sum_{i<j}^{3}\mu_{ij}\left(e_{ij}+e_{ij}^{\dagger}\right)\left(a+a^{\dagger}\right),
\label{eq:2}
\end{multline}

where $J_{z}^{\left(1\right)}=\frac{1}{2}\left(e_{22}-e_{11}\right)$ (half the population difference between the second and first levels), $J_{z}^{\left(2\right)}=\frac{1}{2}\left(e_{33}-e_{22}\right)$ (half the population difference between the third and second levels), $\omega_{1}=-\frac{4}{3}\overline{\omega}_{1}+\frac{2}{3}\overline{\omega}_{2}+\frac{2}{3}\overline{\omega}_{3}$ and $\omega_{2}=-\frac{2}{3}\overline{\omega}_{1}-\frac{2}{3}\overline{\omega}_{2}+\frac{4}{3}\overline{\omega}_{3}$.

\begin{figure}[b]
	\begin{centering}
		\includegraphics[scale=0.33]{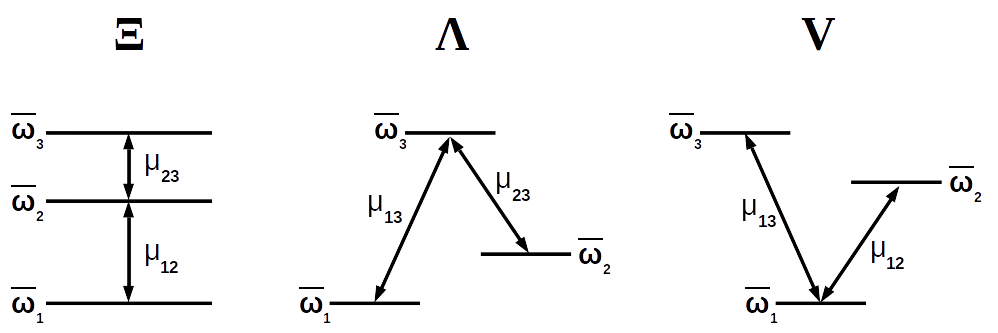}
		\par\end{centering}
	
	\caption{Diagram showing the three possible configurations of a three-level
		atom according to the permitted transitions between its levels.\label{fig:1}}
\end{figure}

Selection rules for a dipolar transition force the parity of the quantum states between which the transition is made to be opposite, and hence to one of the coupling parameters $\mu_{ij}$ to be zero, giving rise to three possible three-level atom configurations: $\Xi$ configuration ($\mu_{13}=0$), $\Lambda$ configuration ($\mu_{12}=0$) and $V$ configuration ($\mu_{23}=0$) (Figure \ref{fig:1}).

\subsubsection*{Representation theory and Cooperation number}

The term cooperation number was first introduced by Dicke in his original paper \cite{key-1}, referring to the different representations of $SU(2)$ used in the description of the full state's space of his hamiltonian. Here we make a brief analysis of the representations of $SU(3)$ and its basis states (Gelfand-Tsetlin states), which we later use to describe the three-level atoms in our system. The influence of the cooperation number over the QPT and some expectation values has already been studied for the Dicke model \cite{key-24}, as well as its effect on the entropy of entanglement in two- and three-level systems \cite{key-24, key-25}.

The operators $J_{z}^{\left(1\right)}$, $J_{z}^{\left(2\right)}$, $e_{12}$, $e_{23}$, $e_{12}^{\dagger}$ and $e_{23}^{\dagger}$ in the hamiltonian (\ref{eq:2}), form a basis for the Lie algebra of $SU(3)$, one that is particularly convenient if it is adopted along with the labeling scheme devised by Gelfand and Tsetlin \cite{key-26} for the basis states of the irreducible representations (irreps) of $SU(n)$. Given an irrep $h=(h_{1},h_{2},h_{3})$ of $SU(3)$, the scheme, called a Gelfand-Tsetlin pattern, is as follows:

\begin{center}
$\left|
\begin{array}{ccc}
h_{1} & h_{2} & h_{3}\\
q_{1} & q_{2}\\
r
\end{array}
\right\rangle$
\end{center}
where the top row contains the information that specifies the irrep, while the entries of lower rows are subject to the betweenness conditions: 
$h_{1}\geq q_{1}\geq h_{2}$, 
$h_{2}\geq q_{2}\geq h_{3}$ and 
$q_{1}\geq r\geq q_{2}$.

These basis states are simultaneous eigenstates of the operators $J_{z}^{\left(1\right)}$ and $J_{z}^{\left(2\right)}$, explicit formulae exist for the matrix elements of $e_{12}$, $e_{23}$, $e_{12}^{\dagger}$ and $e_{23}^{\dagger}$ \cite{key-27} and they allow us to have a very simple physical interpretation of its parameters in our particular context: $r$ is the number of atoms in the first (lowest) energy level, $q_{1}+q_{2}-r$ is equal to the number of atoms in the second energy level and $h_{1}+h_{2}+h_{3}-q_{1}-q_{2}$ is equal to the number of atoms in the third (highest) energy level, where $h_{1}$, $h_{2}$ and $h_{3}$ are subject to the constraint $h_{1}+h_{2}+h_{3}=N$ (the total number of atoms).

The operators of the atomic subsystem in the hamiltonian (\ref{eq:2}) act, in principle, on the complex Hilbert space $\left(\mathbb{C}^{3}\right)^{\otimes N}$, which has a dimension of $3^{N}$; this space can be decomposed into a direct sum of subspaces $\mathcal{H}_{h}$ labeled by the permitted representations $h$ of $SU(3)$ for a given $N$:

\begin{center}
$\left(\mathbb{C}^{3}\right)^{\otimes N}=\ensuremath{\bigoplus\limits _{h}g_{h}\mathcal{H}_{h}}$,
\end{center}
where $g_{h}$ is the representation's multiplicity (the number of times the representation appears in the decomposition) and the sum runs over all possible representations $h=(h_{1},h_{2},h_{3})$ such that $h_{1}+h_{2}+h_{3}=N$ and $h_{1}\geq h_{2}\geq h_{3}$ (from the betweenness condition of the Gelfand-Tsetlin patterns). Nevertheless, working with this space is physically equivalent to studying a system of $N$ fully distinguishable atoms, which we don't usually have in experimental realizations of the studied system. If we were to consider every possible representation with its own multiplicity, we would be treating the atoms as fully distinguishable; on the other hand, if we just consider the symmetric representation ($h_{1}=N$, $h_{2}=h_{3}=0$), we would be treating the atoms as fully indistinguishable. Here we consider every possible representation but ignore its multiplicity, leading us to treat the atoms as semi-distinguishable, the cooperation number being what adds some distinguishability to the states.

The idea behind the term ``cooperation number'', as described by Dicke, is that of an effective number of atoms in the system, however this notion by itself is hard to generalize to $n$-level systems without a proper definition. Here we define the cooperation number ($n_{c}$) to be the maximum difference in the number of atoms between any pair of levels. This number changes depending on the representation of $SU(n)$ we use to describe the system: for an arbitrary representation $h=(h_{1},h_{2},\ldots,h_{n})$ the cooperation number is found to be

\begin{equation}
n_{c}=h_{1}-h_{n}.
\label{eq:3}
\end{equation}
In this particular work, where three-level atoms are being studied, the cooperation number (\ref{eq:3}) is simply $n_{c}=h_{1}-h_{3}$. Notice that a state with $n_{c}=0$ will have an expectation value of $0$ for the energy operator (\ref{eq:2}).

It is worth mentioning that the parameters $h_{1}$, $h_{2}$ and $h_{3}$ are functions which depend on the total number of atoms (a constant) and the eigenvalues of the Casimir operators of $SU(3)$ (which, by definition, commute with the atomic operators and therefore with the hamiltonian (\ref{eq:2})), this means that the representation parameters and hence the cooperation number are constants of motion in the studied model.


\begin{figure*}[t]
	\begin{centering}
		\includegraphics[scale=0.25]{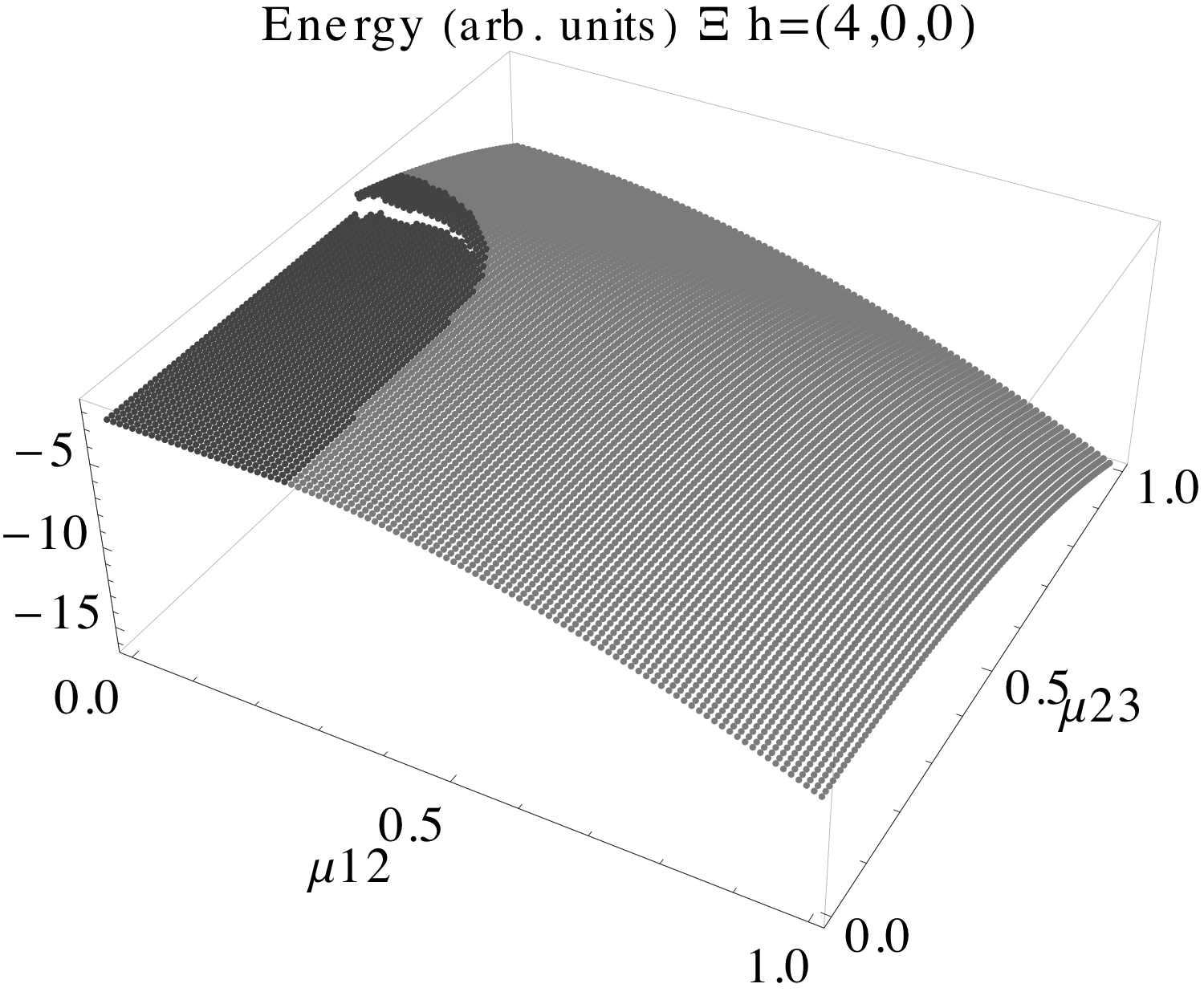}\quad{}\quad{}\quad{}\quad{}\includegraphics[scale=0.25]{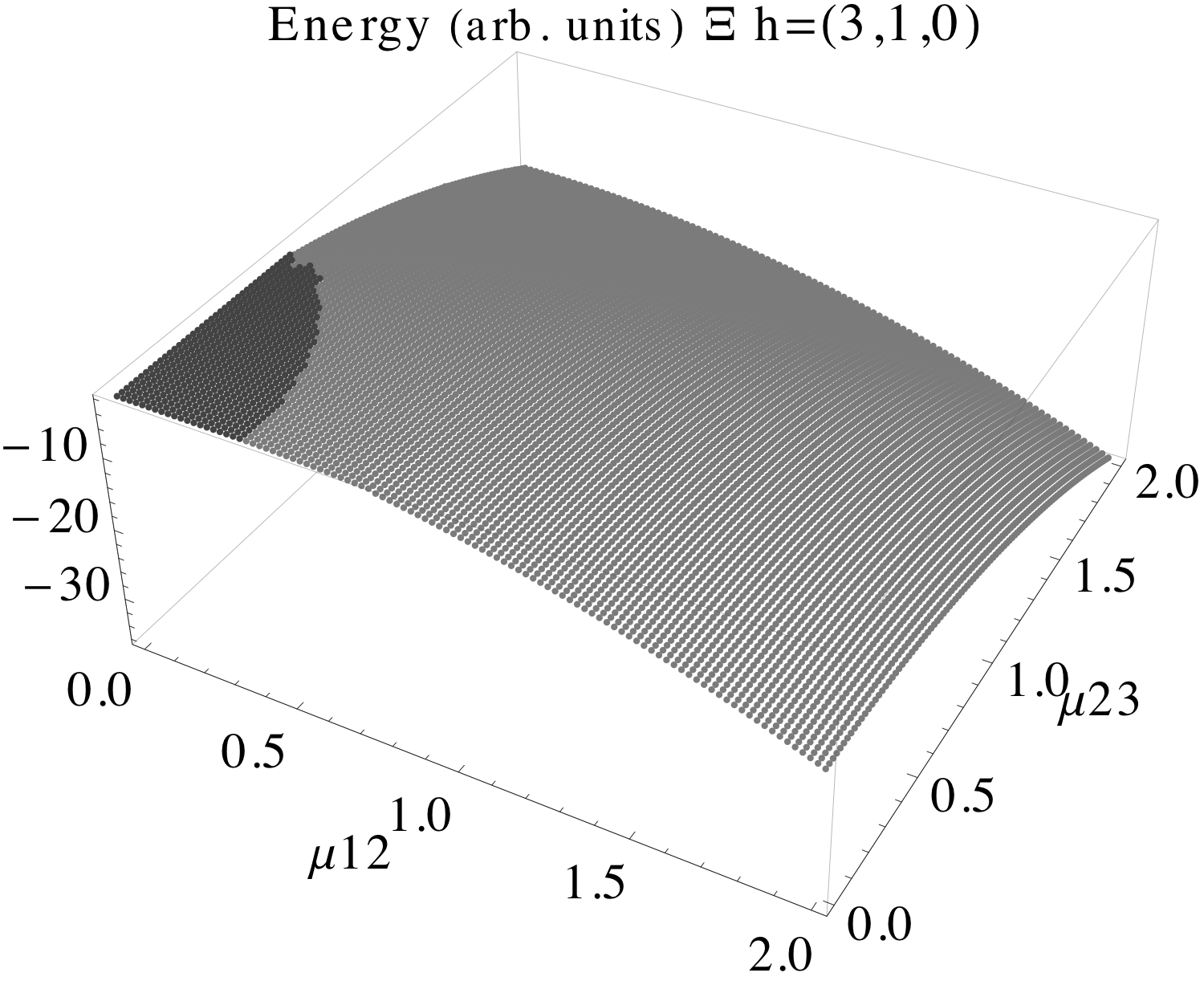}
		\par\end{centering}
	
	\caption{3D plot of the energy of the coherent ground state as a function of
		the coupling parameters $\mu_{12}$ and $\mu_{23}$. The dark-gray
		region represents the normal (sub-radiant) phase of the system. Both
		figures were obtained using $\omega_{1}=1.\bar{3}$, $\omega_{1}=1.\bar{6}$,
		$\Omega=0.5$ and correspond to the $\Xi$ configuration. Left: $h=(4,0,0)$,
		Right: $h=(3,1,0)$. Units are arbitrary but the same for all non-dimensionless
		quantities ($\hbar=1$).\label{fig:2}}
\end{figure*}

\begin{figure*}[t]
	\begin{centering}
		\includegraphics[scale=0.25]{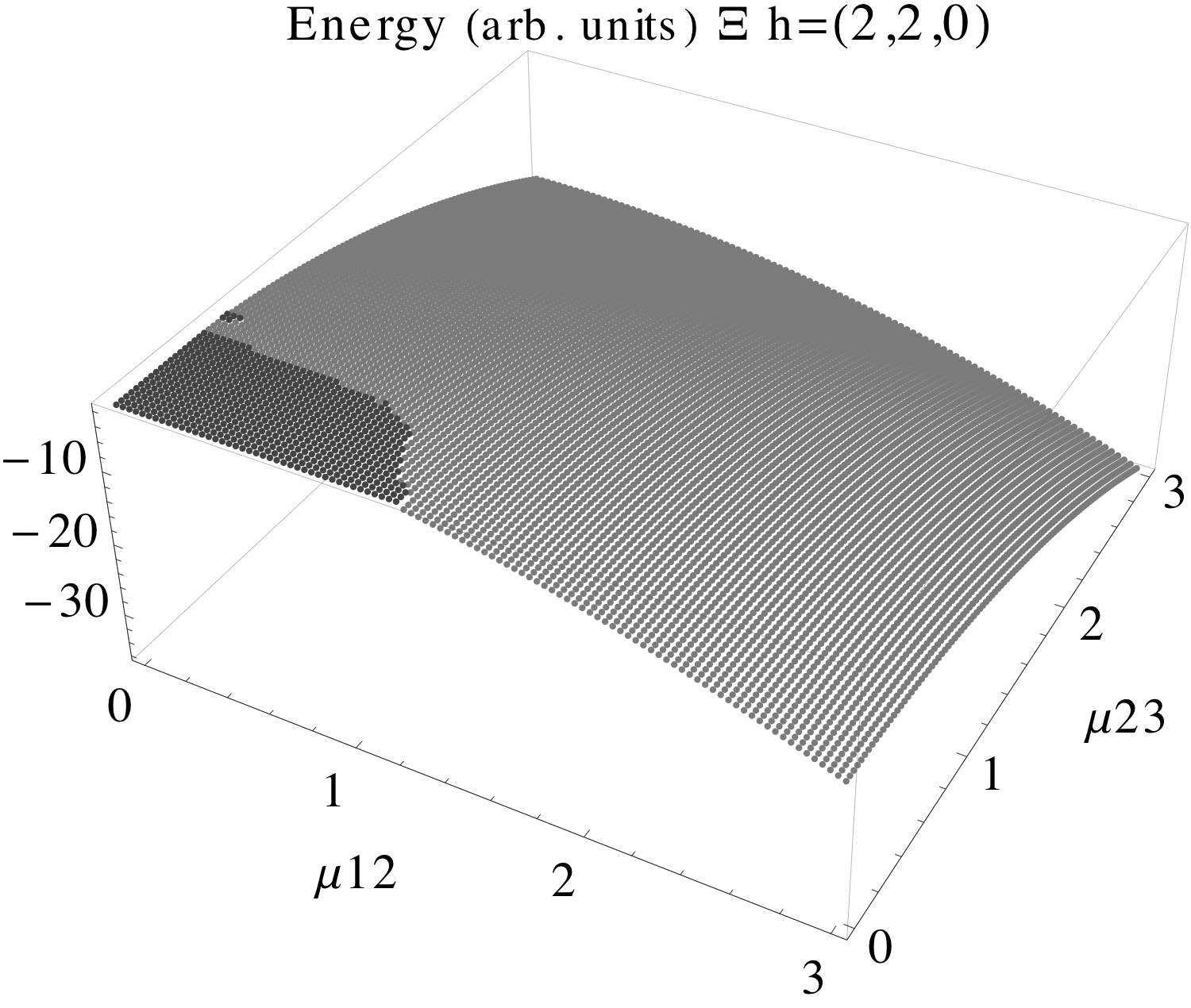}\quad{}\quad{}\quad{}\quad{}\includegraphics[scale=0.25]{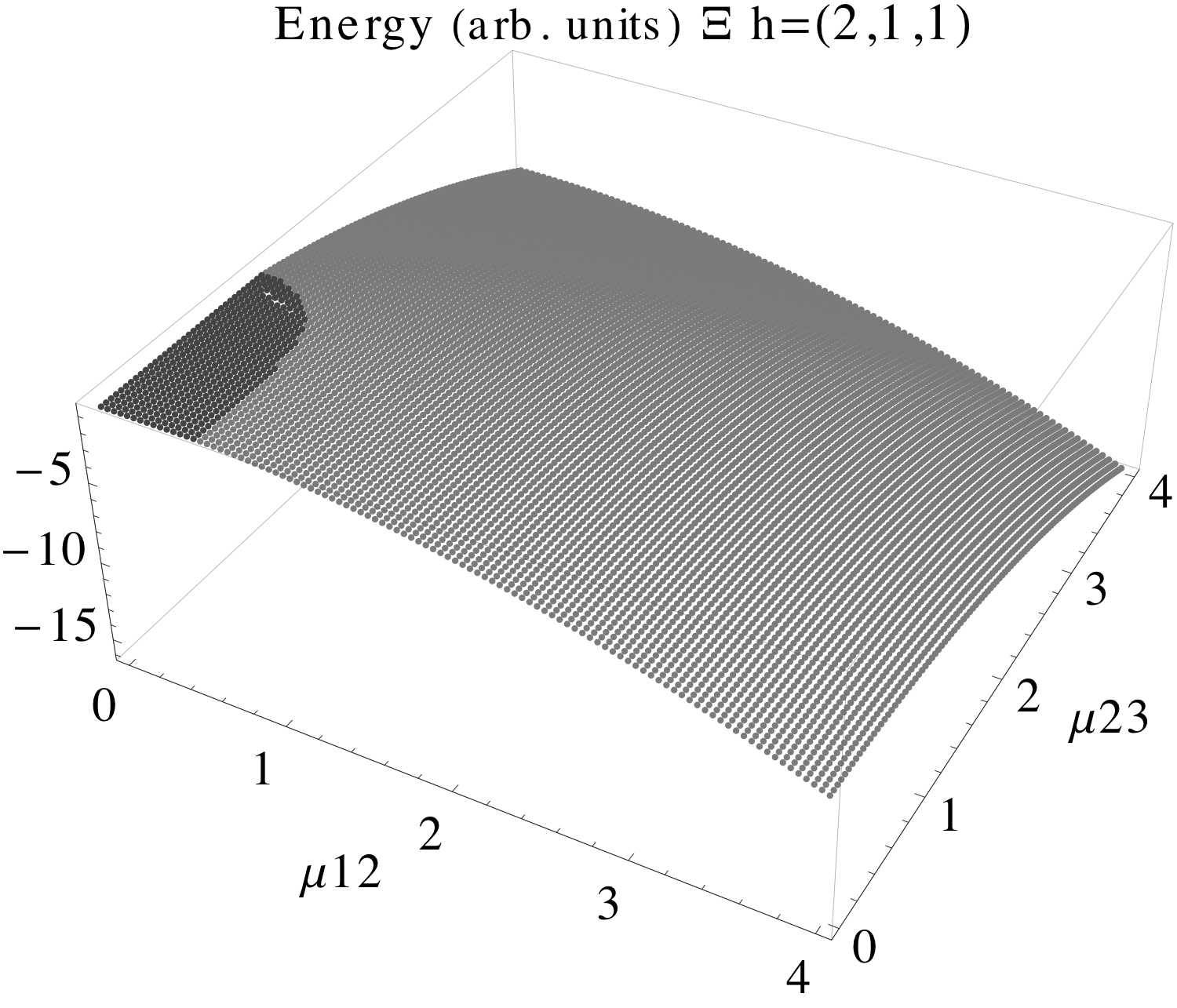}
		\par\end{centering}
	
	\caption{3D plot of the energy of the coherent ground state as a function of
		the coupling parameters $\mu_{12}$ and $\mu_{23}$. The dark-gray
		region represents the normal (sub-radiant) phase of the system. Both
		figures were obtained using $\omega_{1}=1.\bar{3}$, $\omega_{1}=1.\bar{6}$,
		$\Omega=0.5$ and correspond to the $\Xi$ configuration. Left: $h=(2,2,0)$,
		Right: $h=(2,1,1)$. Units are arbitrary but the same for all non-dimensionless
		quantities ($\hbar=1$).\label{fig:3}}
\end{figure*}

\begin{figure*}[t]
	\begin{centering}
		\includegraphics[scale=0.25]{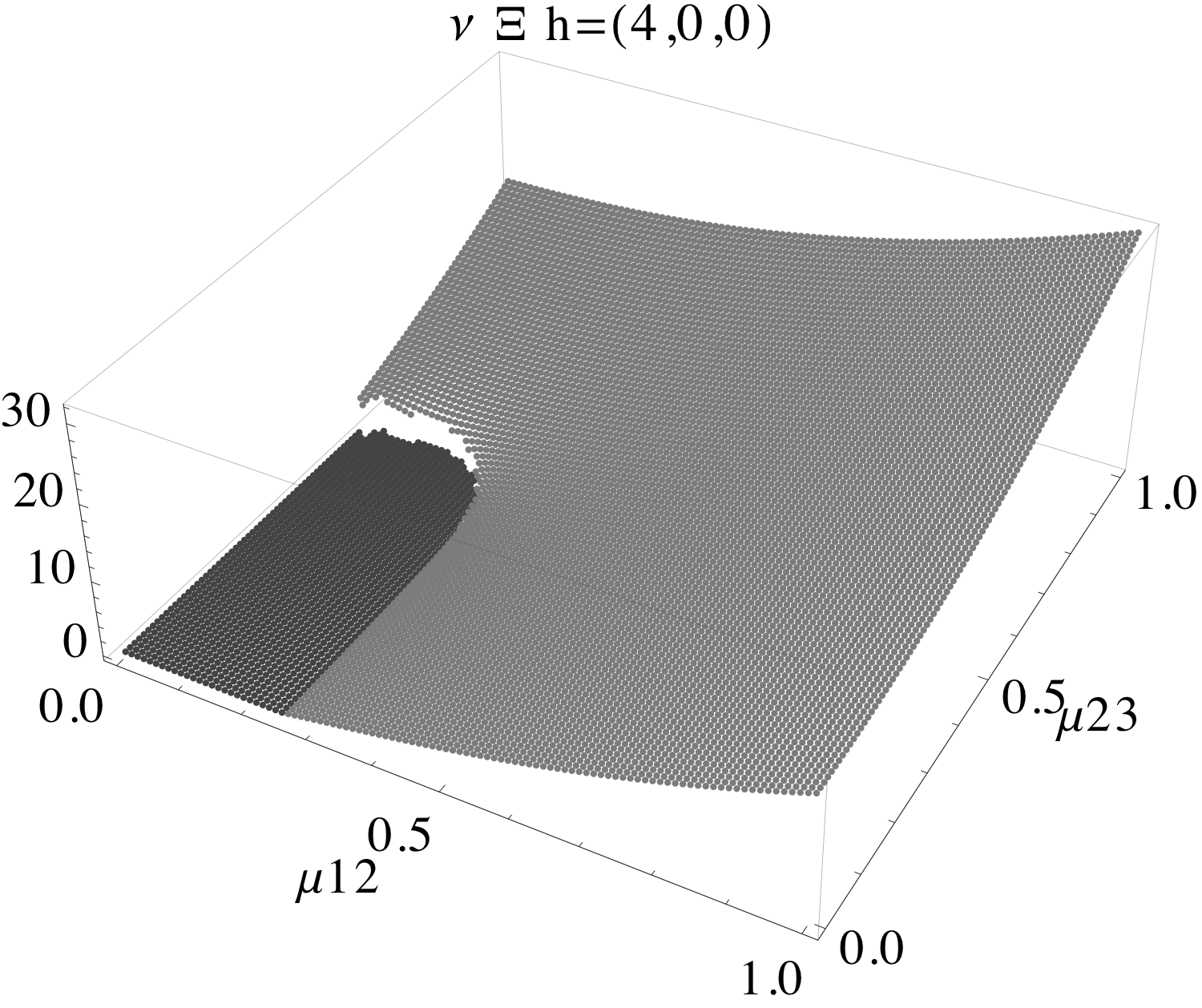}\quad{}\quad{}\quad{}\quad{}\includegraphics[scale=0.25]{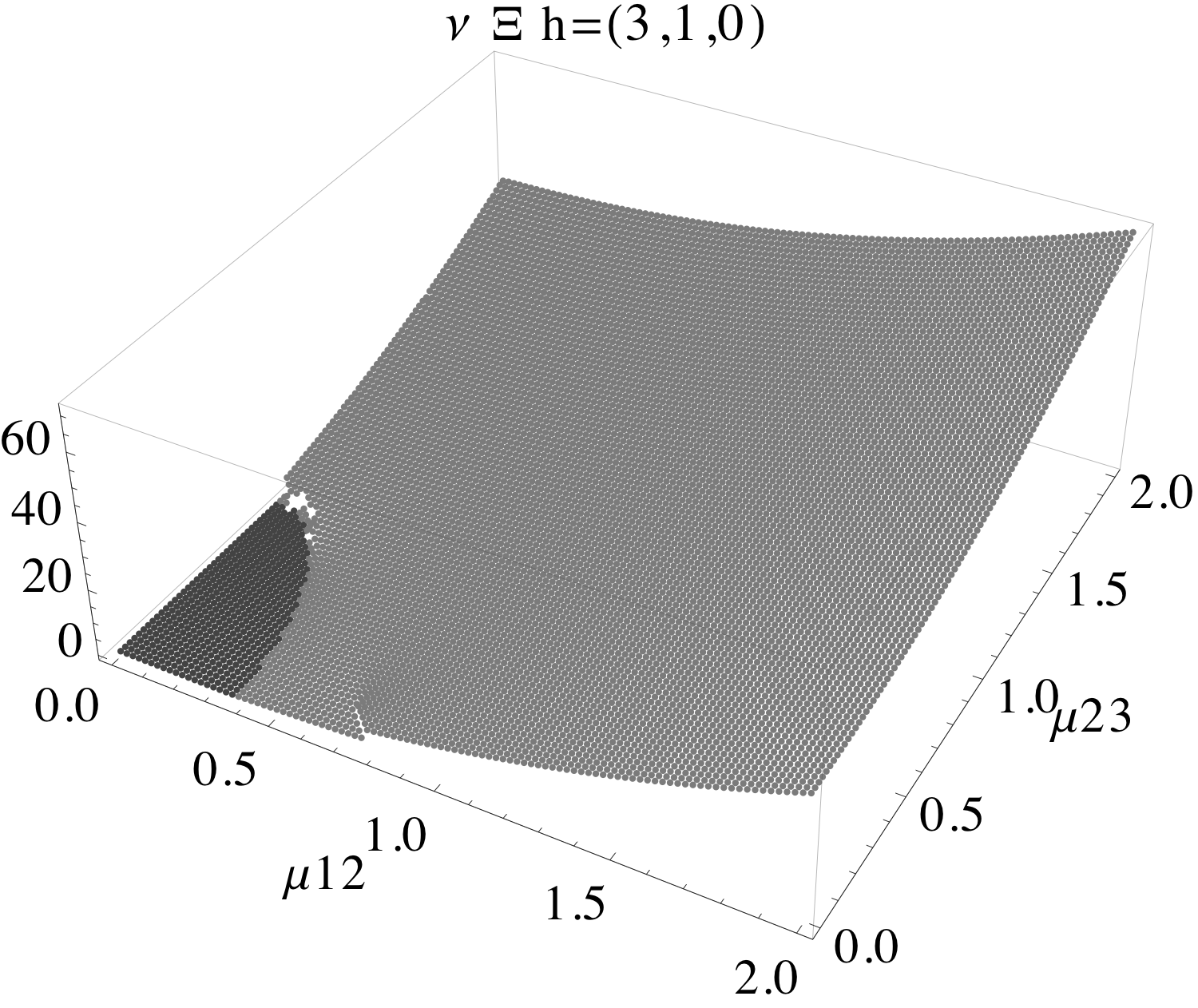}
		\par\end{centering}
	
	\caption{3D plot of the average number of photons in the coherent ground state
		as a function of the coupling parameters $\mu_{12}$ and $\mu_{23}$.
		The dark-gray region represents the normal (sub-radiant) phase of
		the system. Both figures were obtained using $\omega_{1}=1.\bar{3}$,
		$\omega_{1}=1.\bar{6}$, $\Omega=0.5$ and correspond to the $\Xi$
		configuration. Left: $h=(4,0,0)$, Right: $h=(3,1,0)$. Units are
		arbitrary but the same for all non-dimensionless quantities ($\hbar=1$).\label{fig:4}}
\end{figure*}

\begin{figure*}[t]
	\begin{centering}
		\includegraphics[scale=0.25]{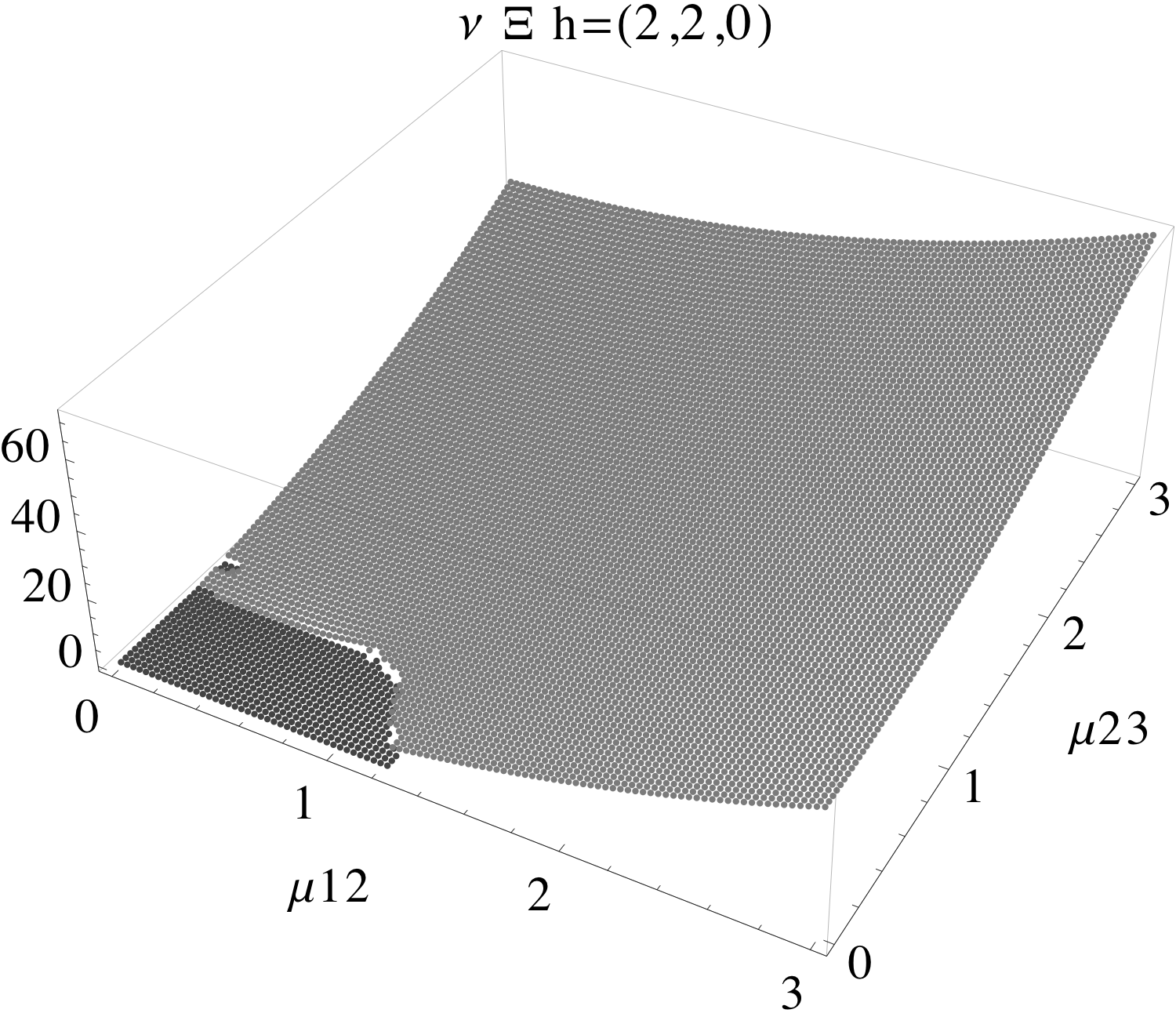}\quad{}\quad{}\quad{}\quad{}\includegraphics[scale=0.25]{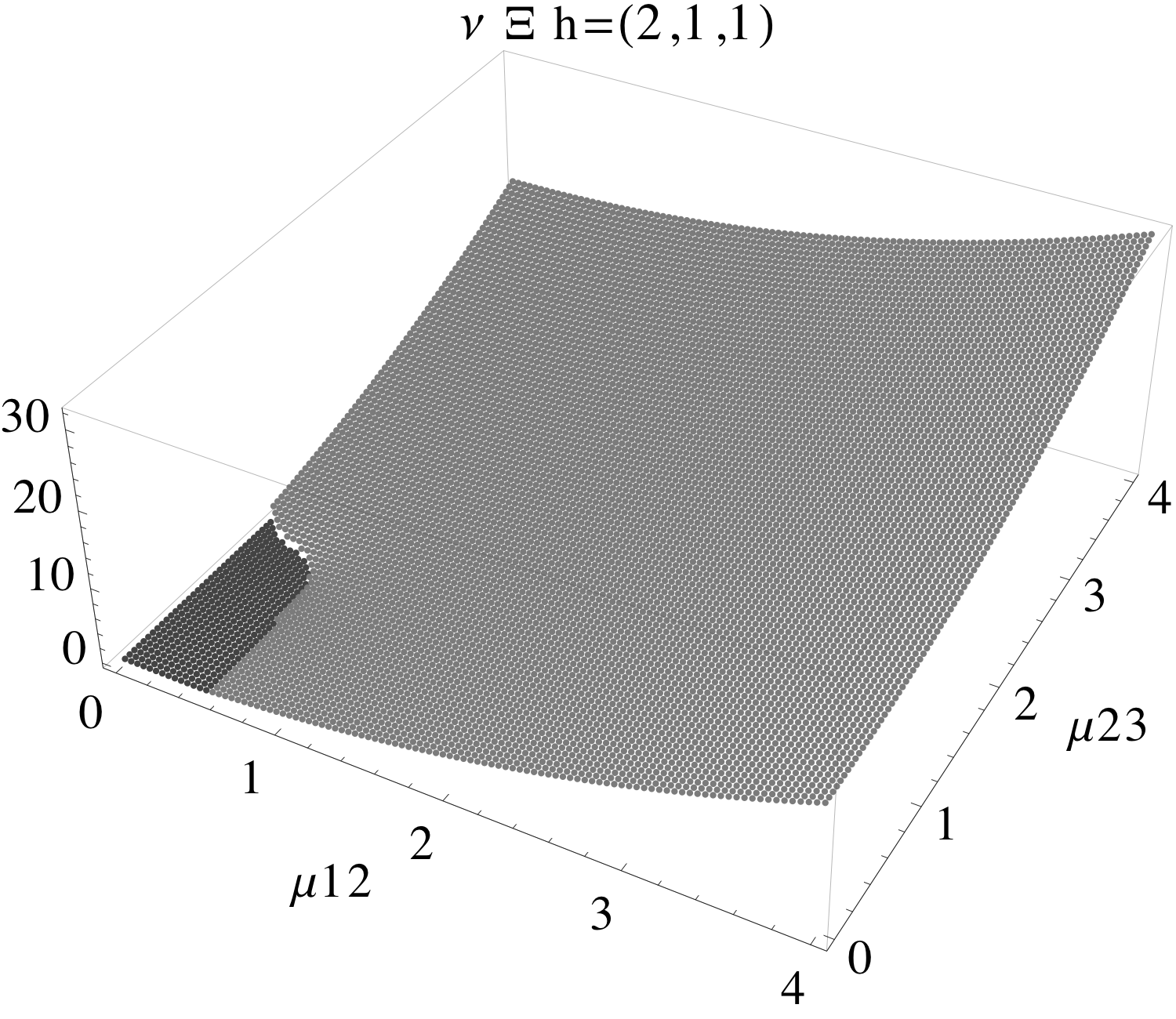}
		\par\end{centering}
	
	\caption{3D plot of the average number of photons in the coherent ground state
		as a function of the coupling parameters $\mu_{12}$ and $\mu_{23}$.
		The dark-gray region represents the normal (sub-radiant) phase of
		the system. Both figures were obtained using $\omega_{1}=1.\bar{3}$,
		$\omega_{1}=1.\bar{6}$, $\Omega=0.5$ and correspond to the $\Xi$
		configuration. Left: $h=(2,2,0)$, Right: $h=(2,1,1)$. Units are
		arbitrary but the same for all non-dimensionless quantities ($\hbar=1$).\label{fig:5}}
\end{figure*}

\begin{figure*}[t]
	\begin{centering}
		\includegraphics[scale=0.25]{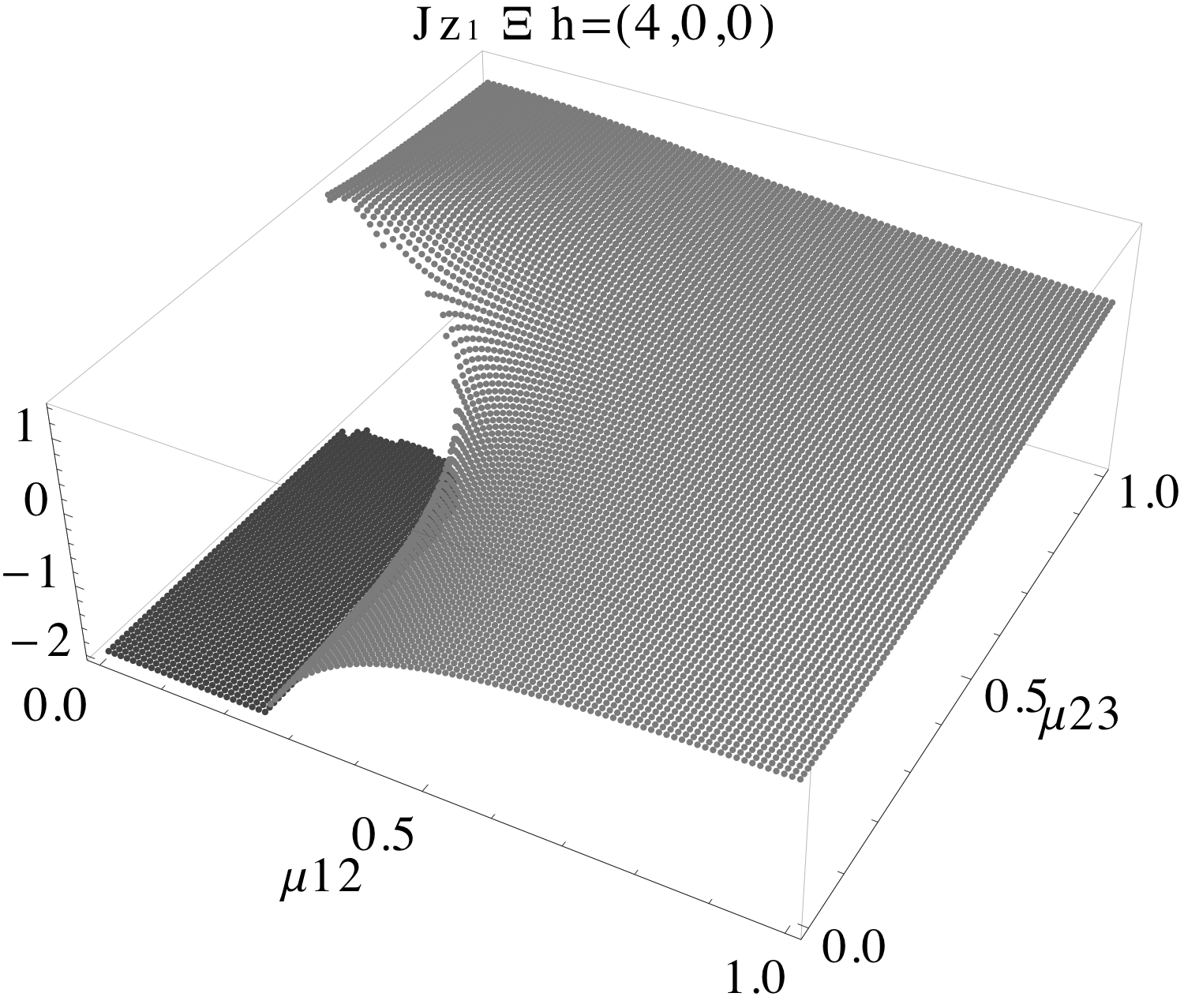}\quad{}\quad{}\quad{}\quad{}\includegraphics[scale=0.25]{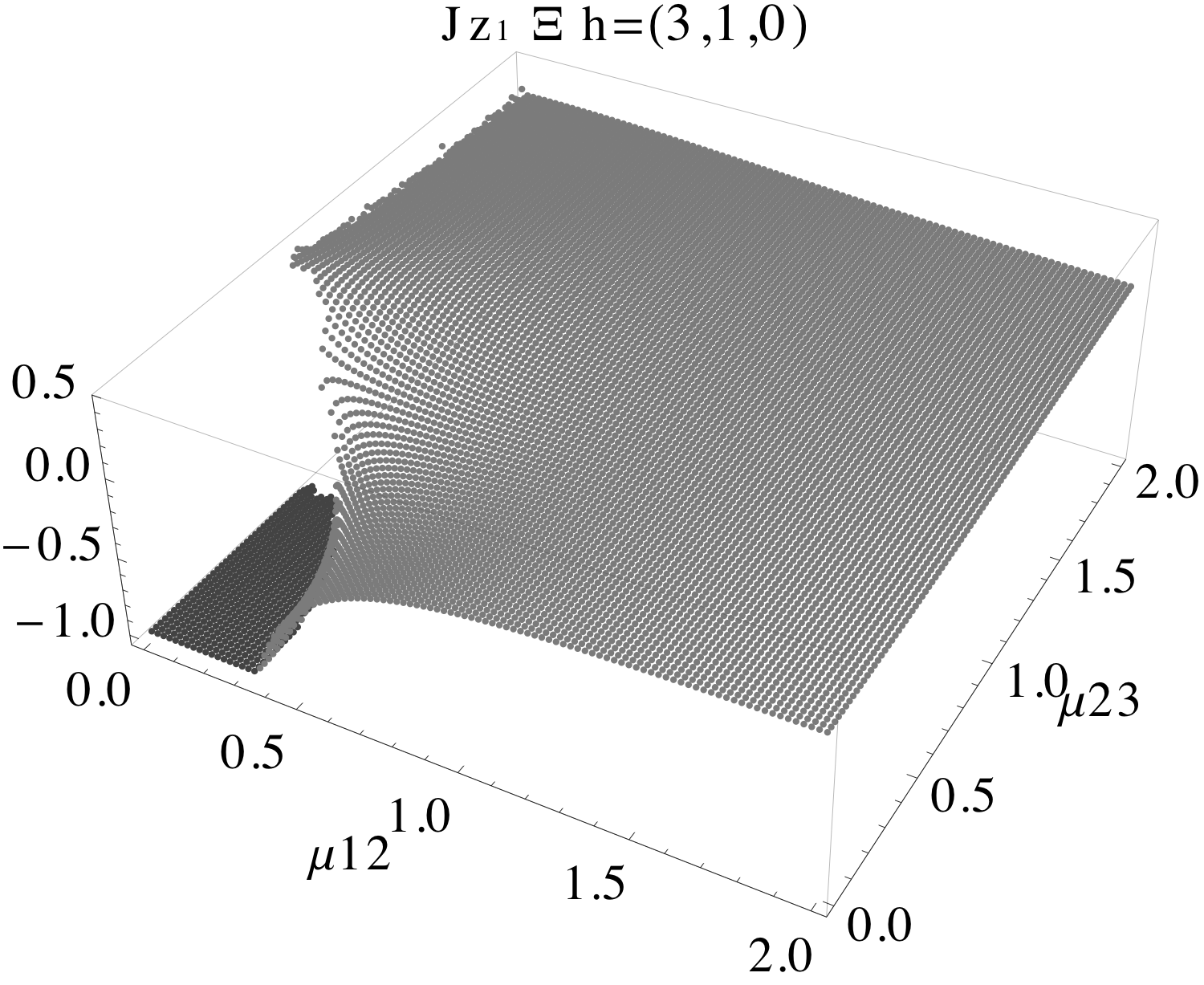}
		\par\end{centering}
	
	\caption{3D plot of the expectation value of the $Jz_{1}$ operator (half the
		population difference between the second and first levels) in the
		coherent ground state as a function of the coupling parameters $\mu_{12}$
		and $\mu_{23}$. The dark-gray region represents the normal (sub-radiant)
		phase of the system. Both figures were obtained using $\omega_{1}=1.\bar{3}$,
		$\omega_{1}=1.\bar{6}$, $\Omega=0.5$ and correspond to the $\Xi$
		configuration. Left: $h=(4,0,0)$, Right: $h=(3,1,0)$. Units are
		arbitrary but the same for all non-dimensionless quantities ($\hbar=1$).\label{fig:6}}
\end{figure*}

\begin{figure*}[t]
	\begin{centering}
		\includegraphics[scale=0.25]{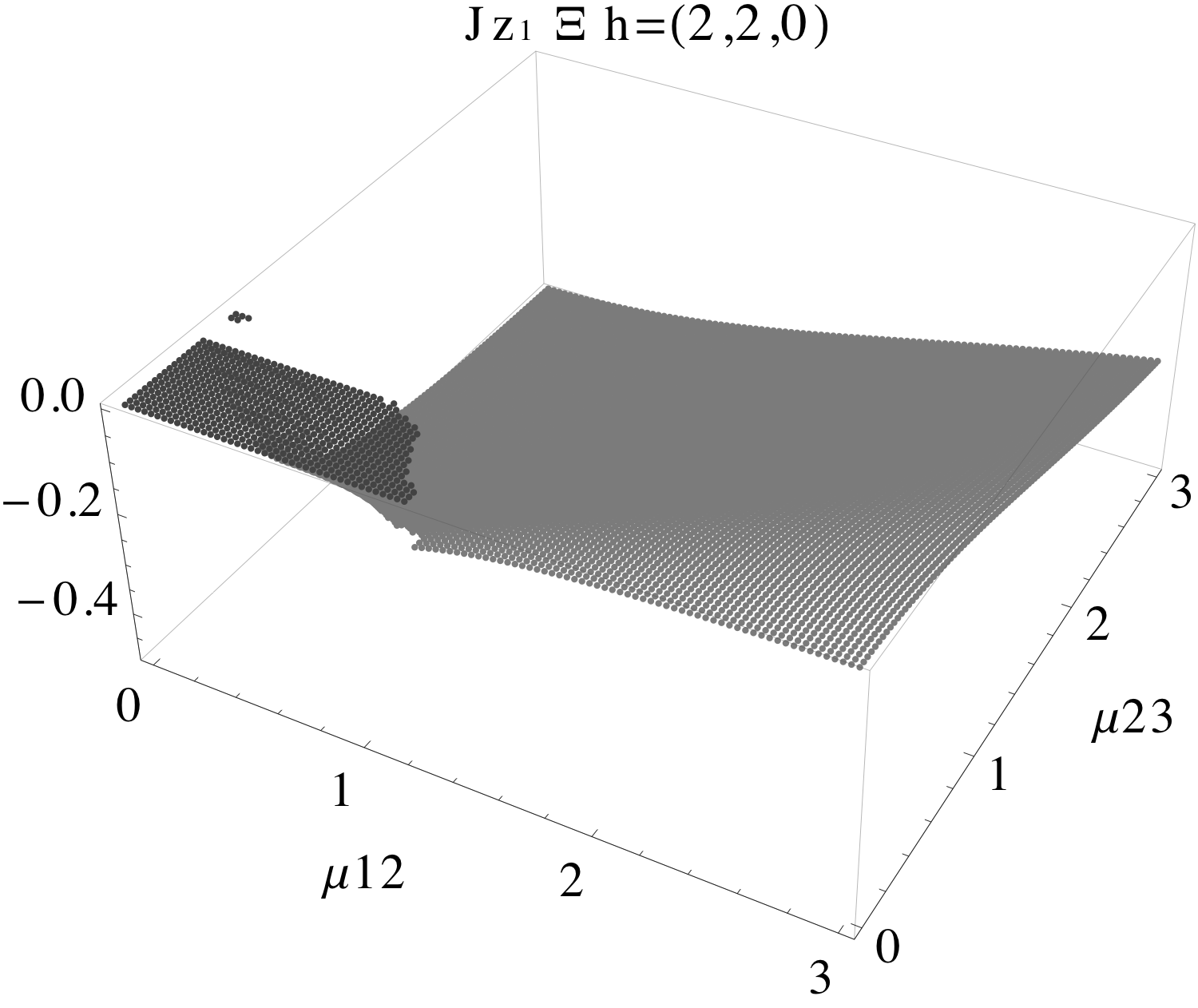}\quad{}\quad{}\quad{}\quad{}\includegraphics[scale=0.25]{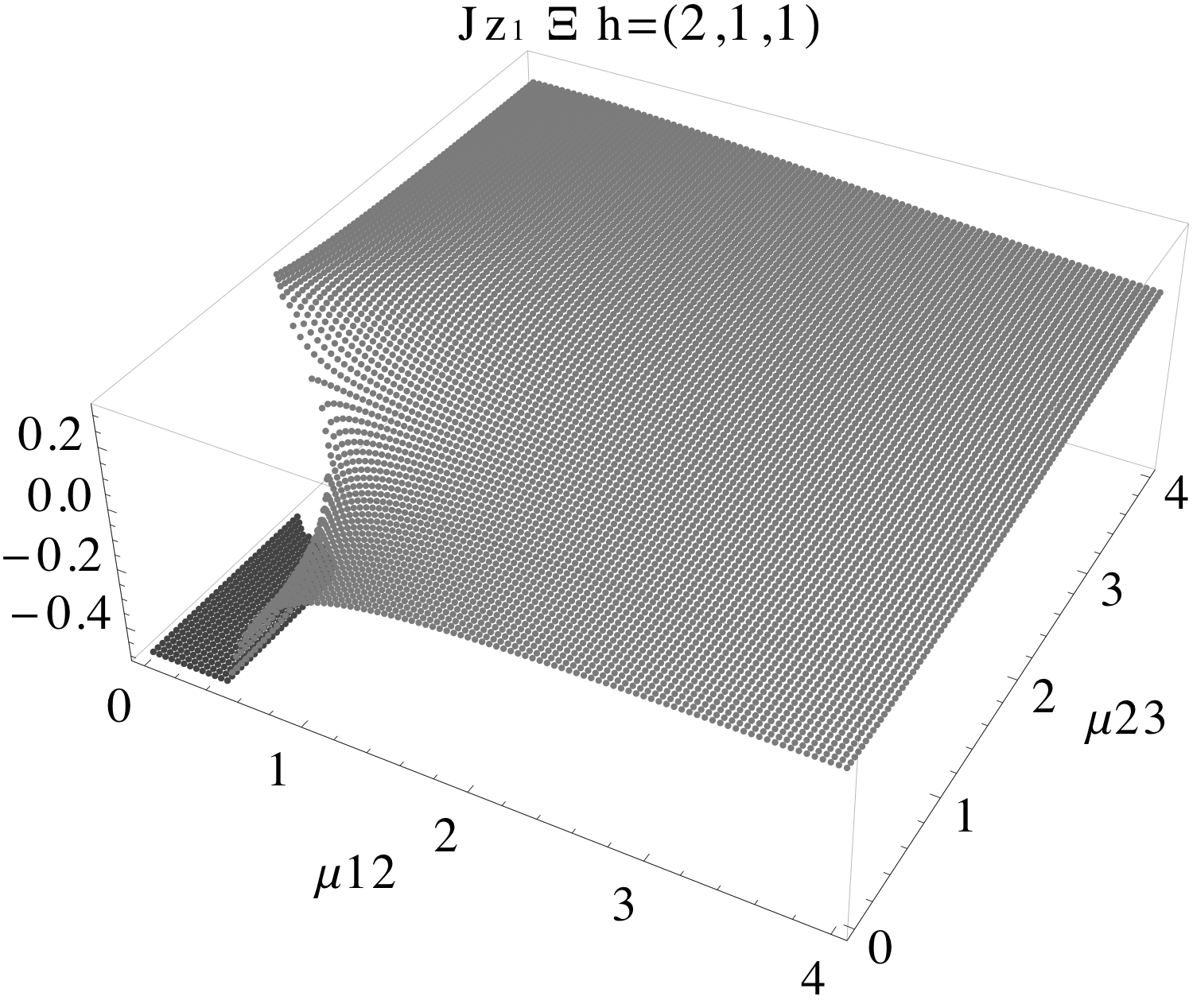}
		\par\end{centering}
	
	\caption{3D plot of the expectation value of the $Jz_{1}$ operator (half the
		population difference between the second and first levels) in the
		coherent ground state as a function of the coupling parameters $\mu_{12}$
		and $\mu_{23}$. The dark-gray region represents the normal (sub-radiant)
		phase of the system. Both figures were obtained using $\omega_{1}=1.\bar{3}$,
		$\omega_{1}=1.\bar{6}$, $\Omega=0.5$ and correspond to the $\Xi$
		configuration. Left: $h=(2,2,0)$, Right: $h=(2,1,1)$. Units are
		arbitrary but the same for all non-dimensionless quantities ($\hbar=1$).\label{fig:7}}
\end{figure*}

\begin{figure*}[t]
	\begin{centering}
		\includegraphics[scale=0.25]{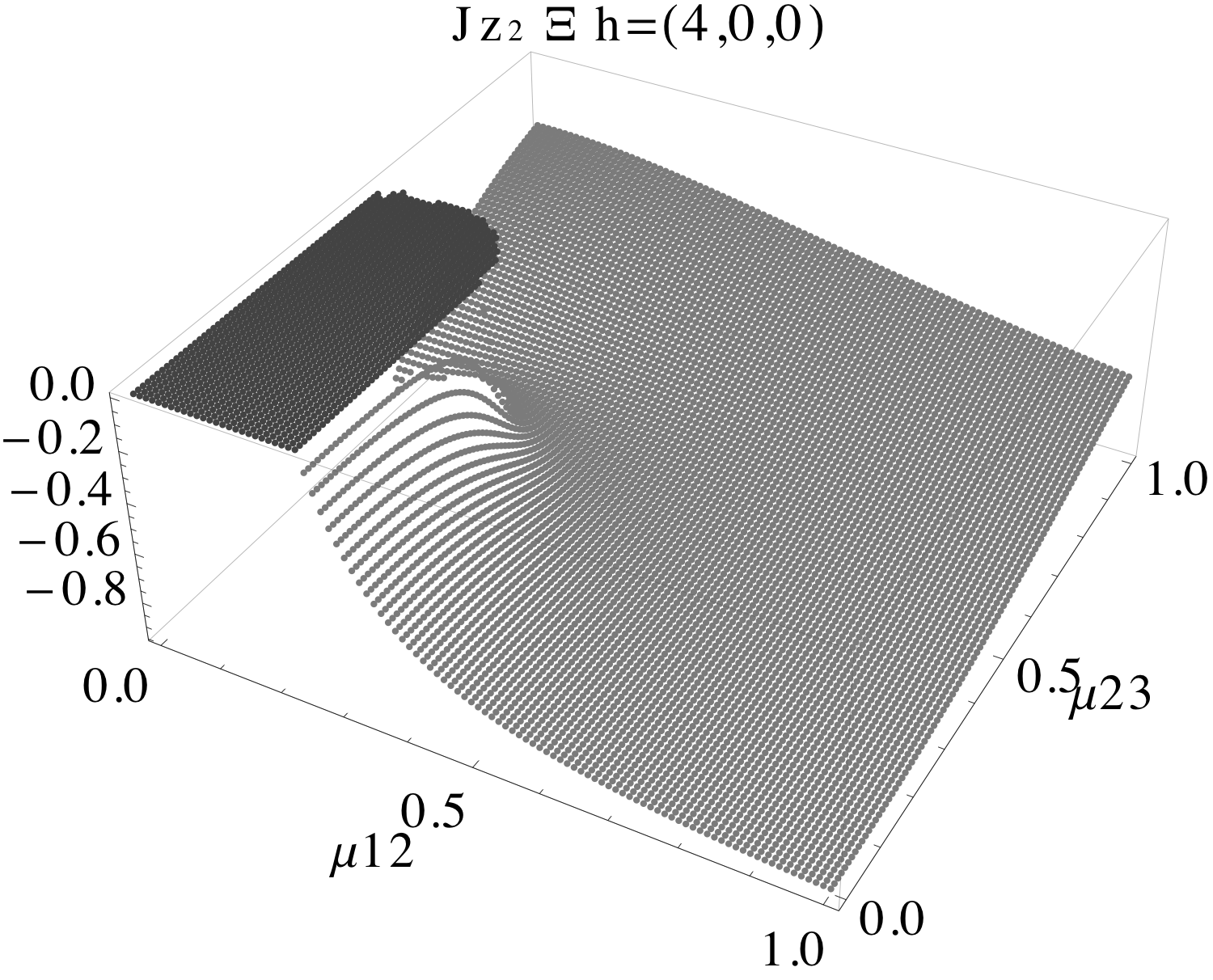}\quad{}\quad{}\quad{}\quad{}\includegraphics[scale=0.25]{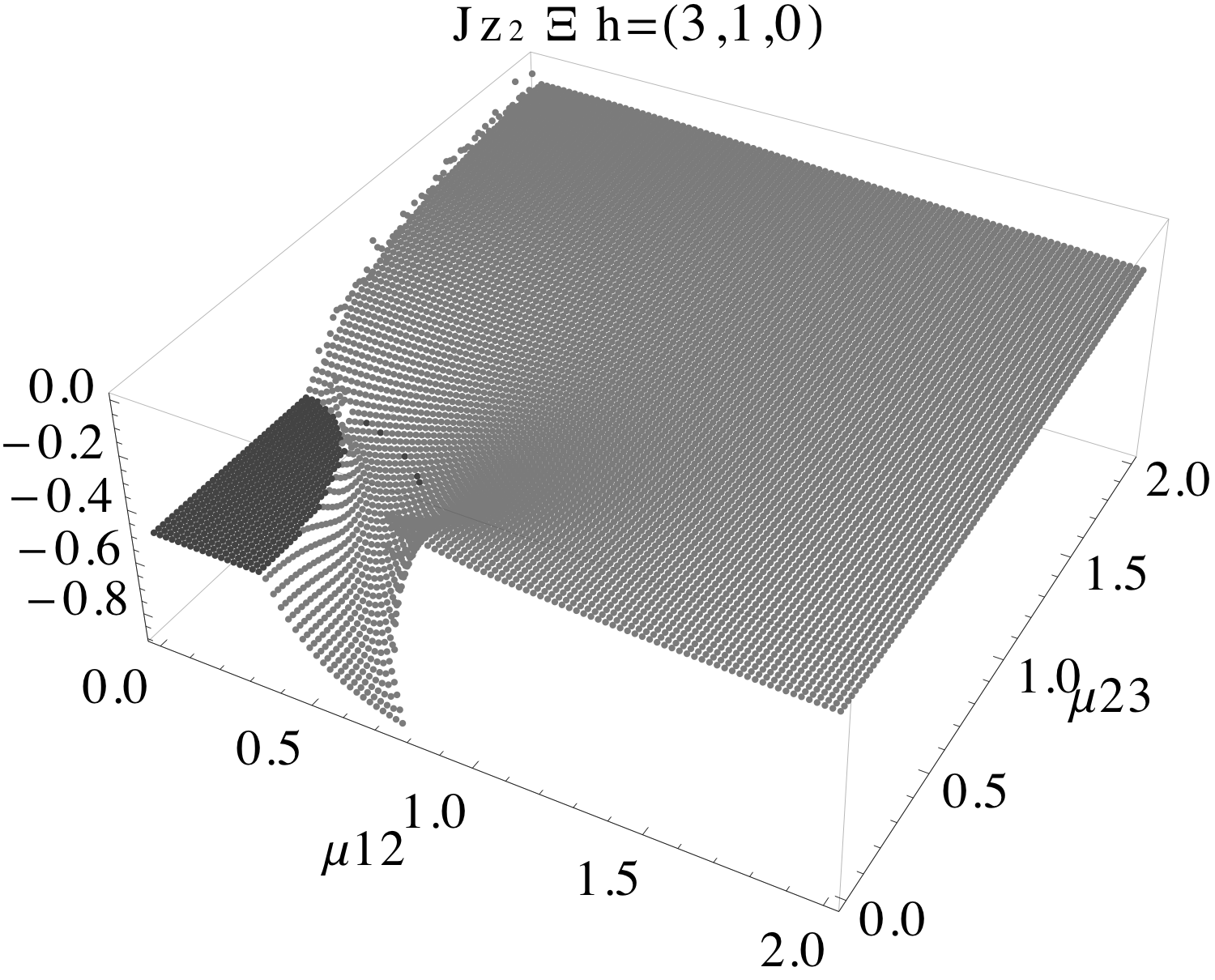}
		\par\end{centering}
	
	\caption{3D plot of the expectation value of the $Jz_{2}$ operator (half the
		population difference between the third and second levels) in the
		coherent ground state as a function of the coupling parameters $\mu_{12}$
		and $\mu_{23}$. The dark-gray region represents the normal (sub-radiant)
		phase of the system. Both figures were obtained using $\omega_{1}=1.\bar{3}$,
		$\omega_{1}=1.\bar{6}$, $\Omega=0.5$ and correspond to the $\Xi$
		configuration. Left: $h=(4,0,0)$, Right: $h=(3,1,0)$. Units are
		arbitrary but the same for all non-dimensionless quantities ($\hbar=1$).\label{fig:8}}
\end{figure*}

\begin{figure*}[t]
	\begin{centering}
		\includegraphics[scale=0.25]{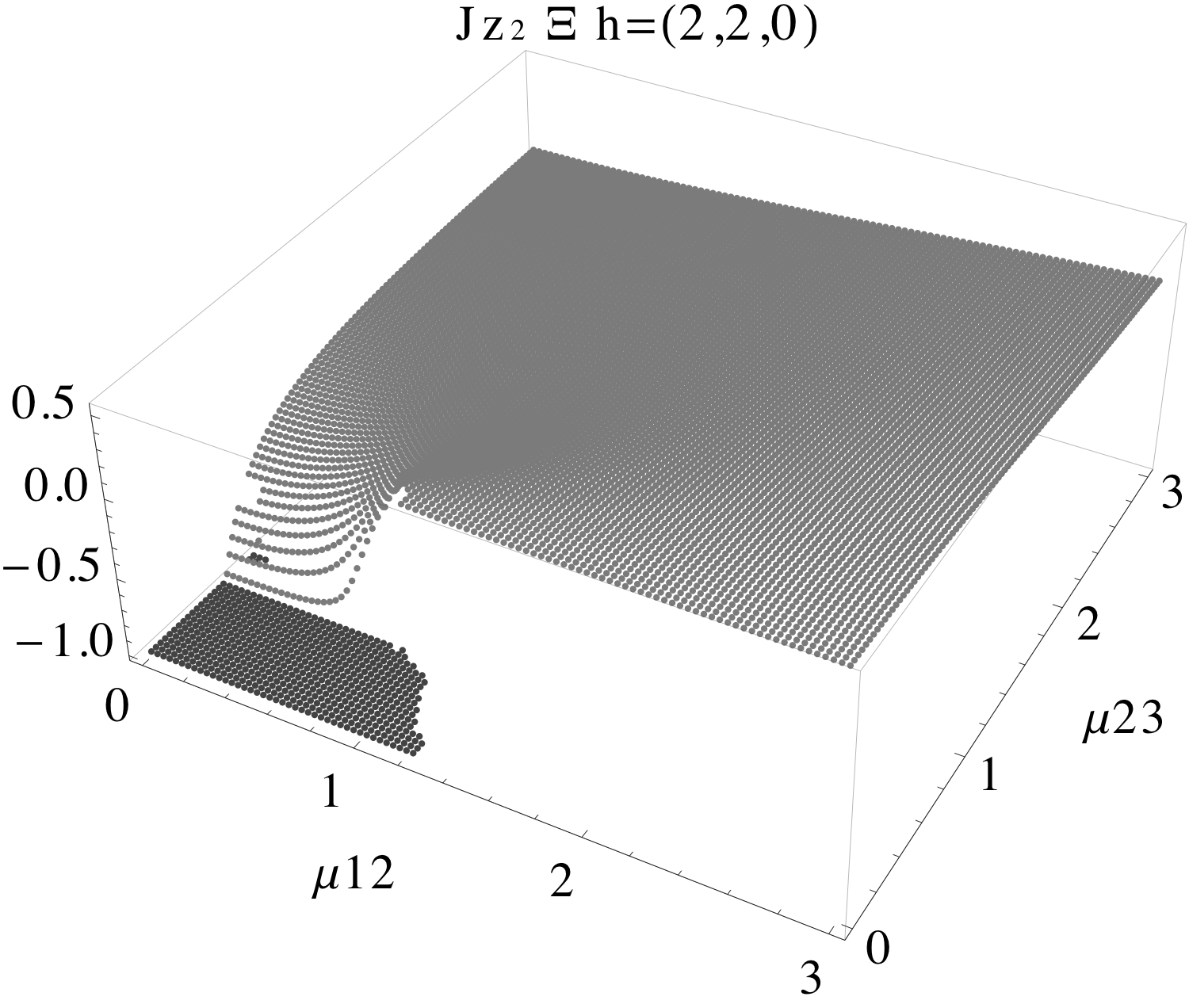}\quad{}\quad{}\quad{}\quad{}\includegraphics[scale=0.25]{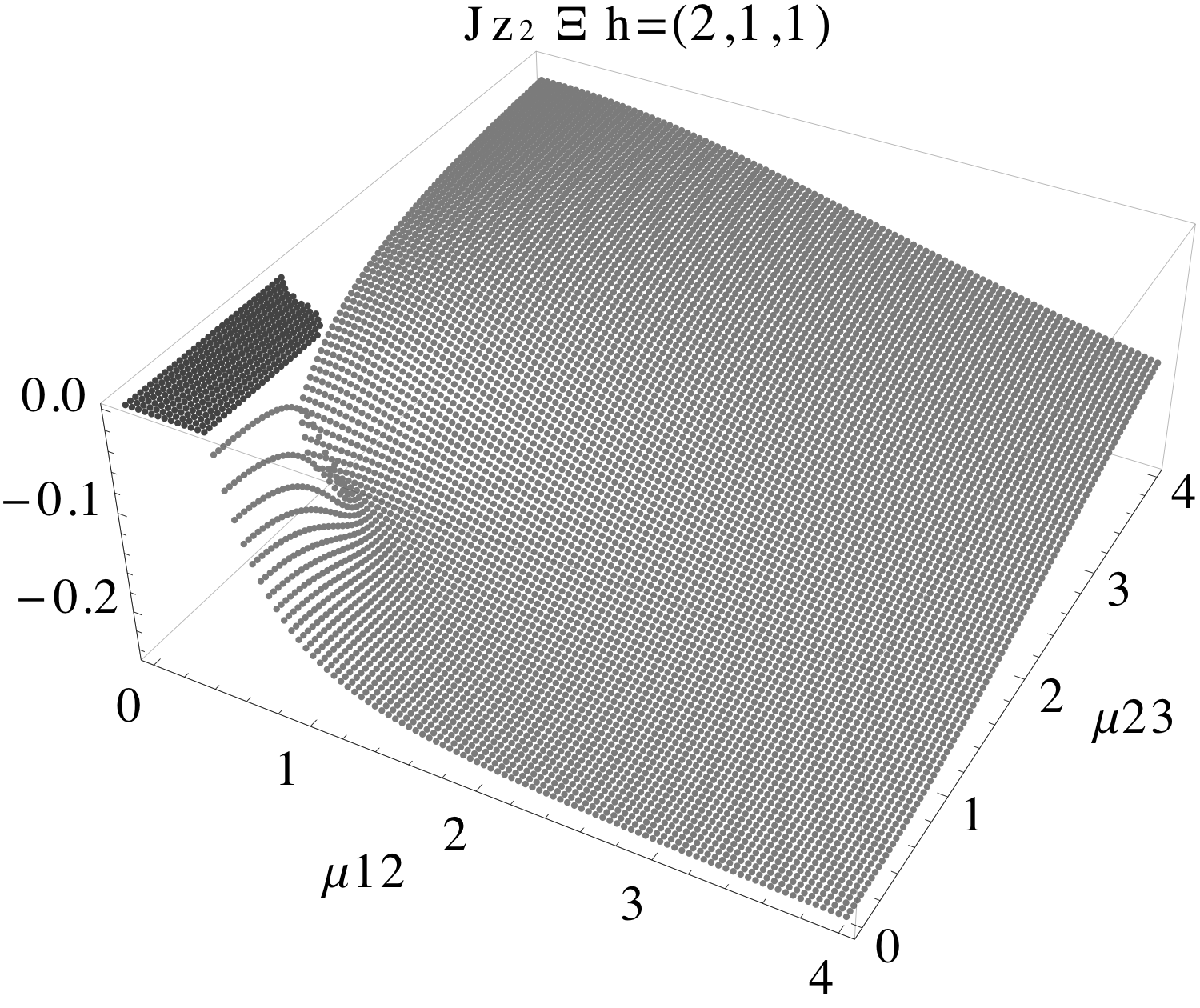}
		\par\end{centering}
	
	\caption{3D plot of the expectation value of the $Jz_{2}$ operator (half the
		population difference between the third and second levels) in the
		coherent ground state as a function of the coupling parameters $\mu_{12}$
		and $\mu_{23}$. The dark-gray region represents the normal (sub-radiant)
		phase of the system. Both figures were obtained using $\omega_{1}=1.\bar{3}$,
		$\omega_{1}=1.\bar{6}$, $\Omega=0.5$ and correspond to the $\Xi$
		configuration. Left: $h=(2,2,0)$, Right: $h=(2,1,1)$. Units are
		arbitrary but the same for all non-dimensionless quantities ($\hbar=1$).\label{fig:9}}
\end{figure*}

\begin{figure*}[t]
	\begin{centering}
		\includegraphics[scale=0.4]{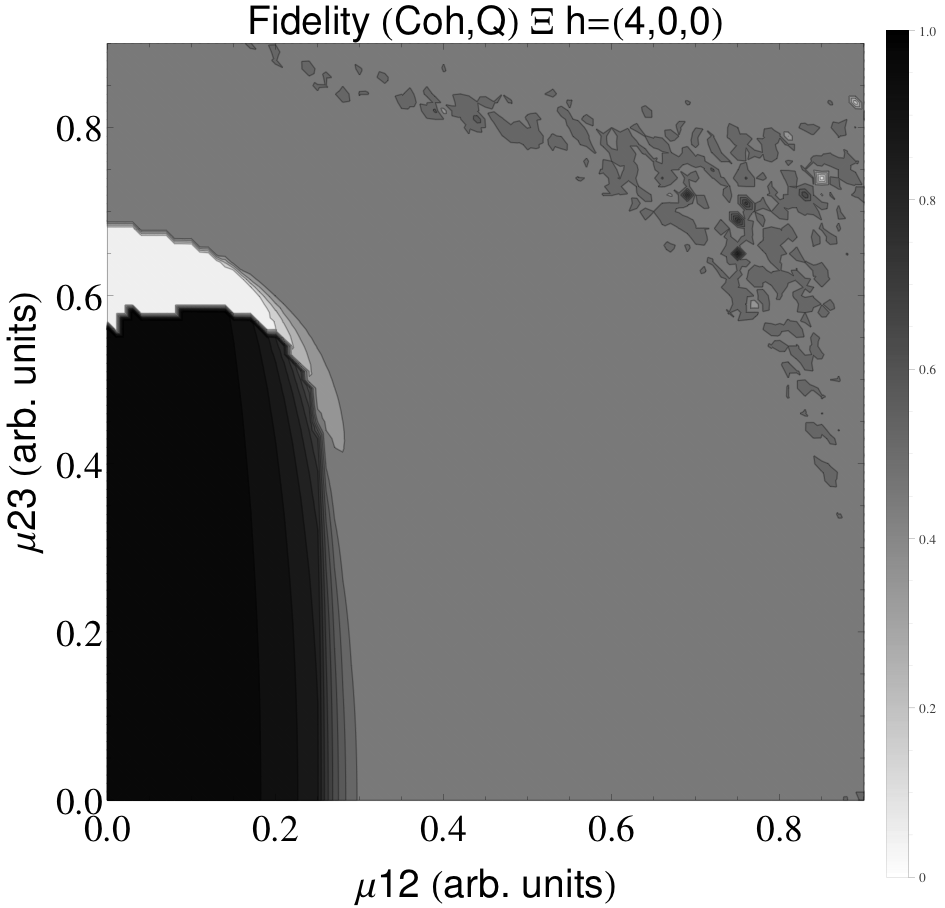}\quad{}\quad{}\quad{}\quad{}\includegraphics[scale=0.33]{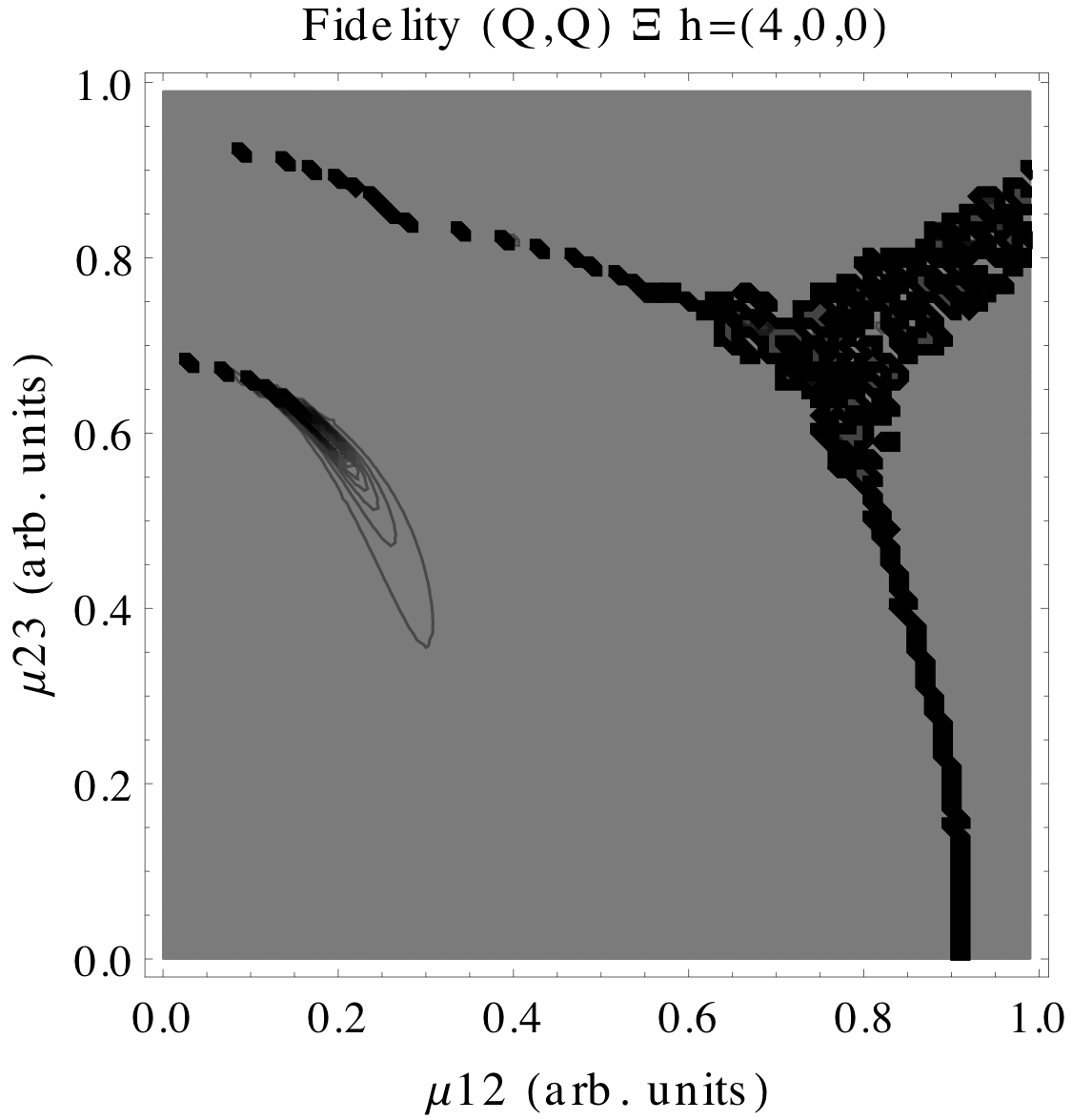}\quad{}
		\par\end{centering}
	
	\caption{Left: Contour plot of the fidelity between coherent states and quantum
		solution as a function of the coupling parameters $\mu_{12}$ and
		$\mu_{23}$, values range between 0 (white) and 1 (black). Right:
		Contour plot of the fidelity between neighboring quantum states as
		a function of the coupling parameters $\mu_{12}$ and $\mu_{23}$,
		black dots represent a drop in the fidelity below 1. Both figures
		were obtained using $\omega_{1}=1.\bar{3}$, $\omega_{1}=1.\bar{6}$,
		$\Omega=0.5$ and correspond to the $\Xi$ configuration in the $h=(4,0,0)$
		representation. Units are arbitrary but the same for all non-dimensionless
		quantities ($\hbar=1$). (Noise in the plots is due to numerical minimization; see text.)\label{fig:10}}
\end{figure*}

\begin{figure*}[t]
	\begin{centering}
		\includegraphics[scale=0.39]{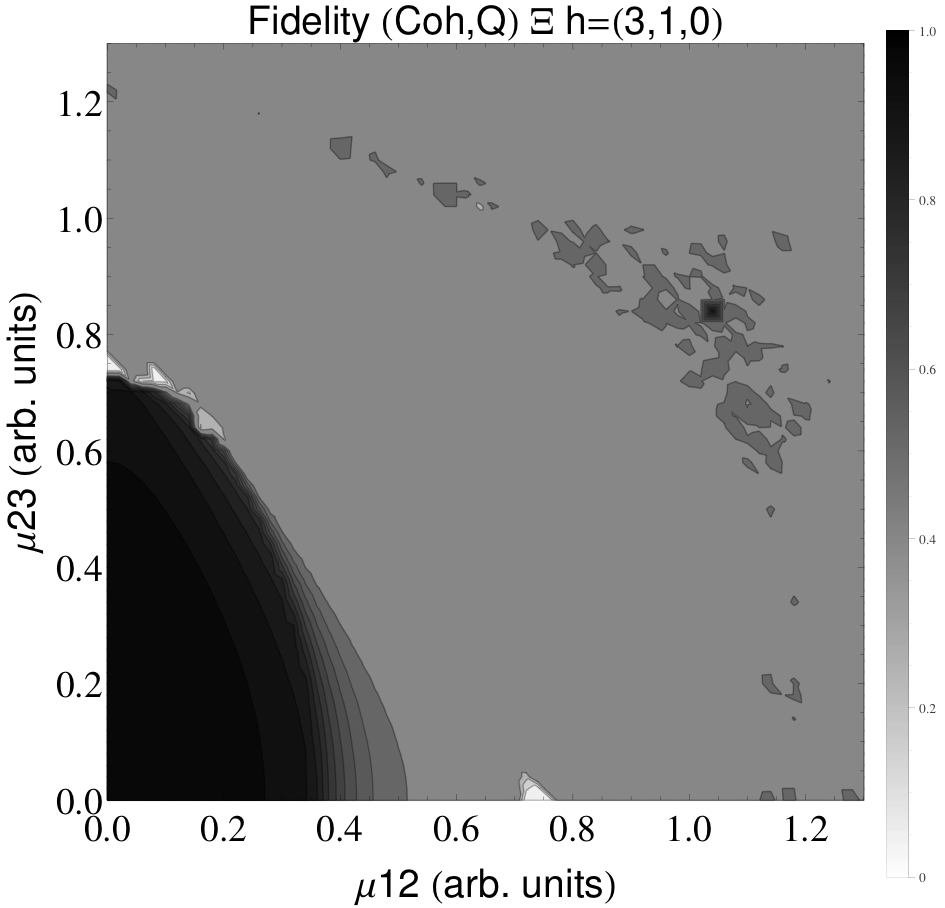}\quad{}\quad{}\quad{}\quad{}\includegraphics[scale=0.24]{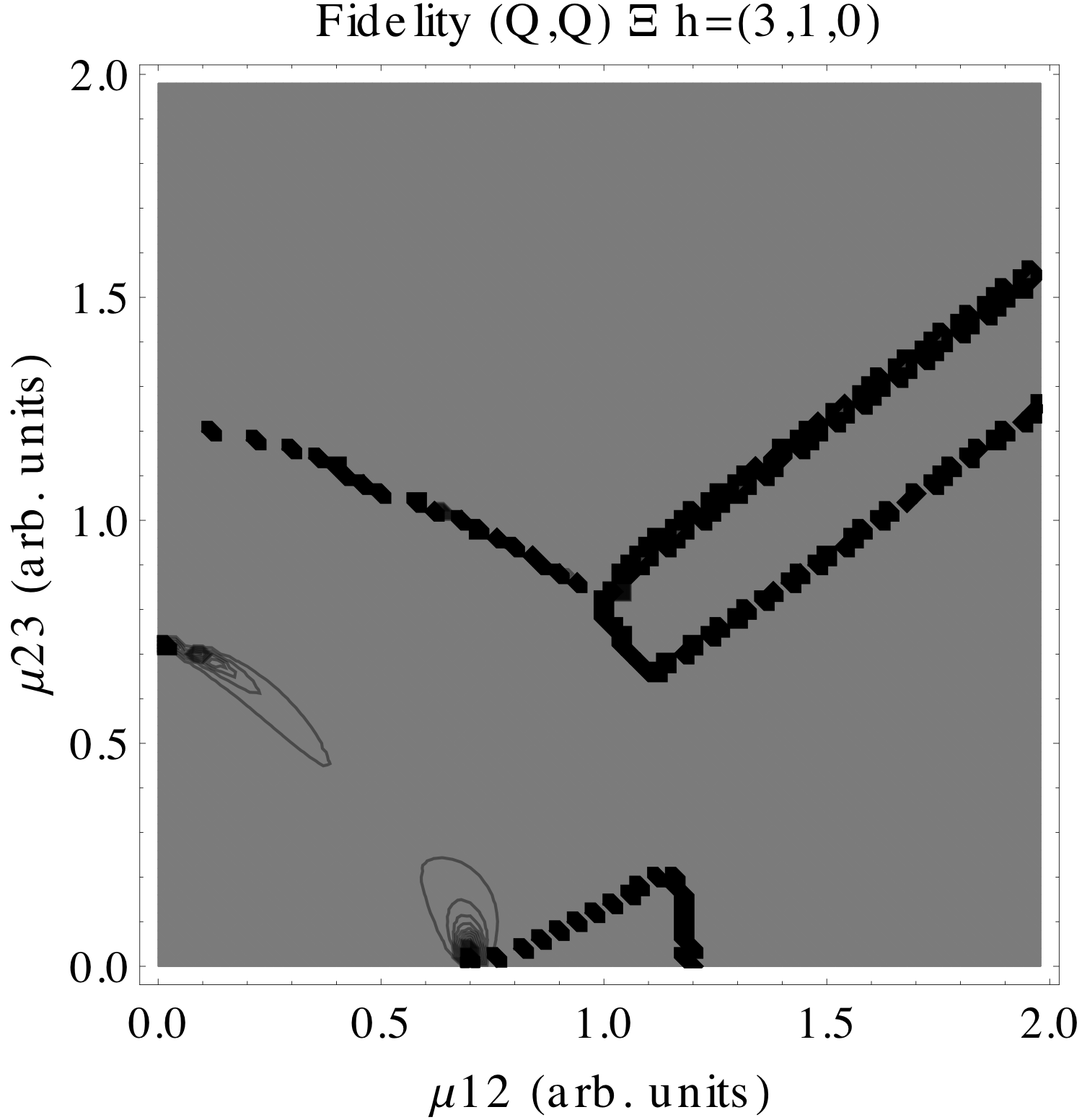}\quad{}
		\par\end{centering}
	
	\caption{Left: Contour plot of the fidelity between coherent states and quantum
		solution as a function of the coupling parameters $\mu_{12}$ and
		$\mu_{23}$, values range between 0 (white) and 1 (black). Right:
		Contour plot of the fidelity between neighboring quantum states as
		a function of the coupling parameters $\mu_{12}$ and $\mu_{23}$,
		black dots represent a drop in the fidelity below 1. Both figures
		were obtained using $\omega_{1}=1.\bar{3}$, $\omega_{1}=1.\bar{6}$,
		$\Omega=0.5$ and correspond to the $\Xi$ configuration in the $h=(3,1,0)$
		representation. Units are arbitrary but the same for all non-dimensionless
		quantities ($\hbar=1$). (Noise in the plots is due to numerical minimization; see text.)\label{fig:11}}
\end{figure*}

\begin{figure*}[t]
	\begin{centering}
		\includegraphics[scale=0.39]{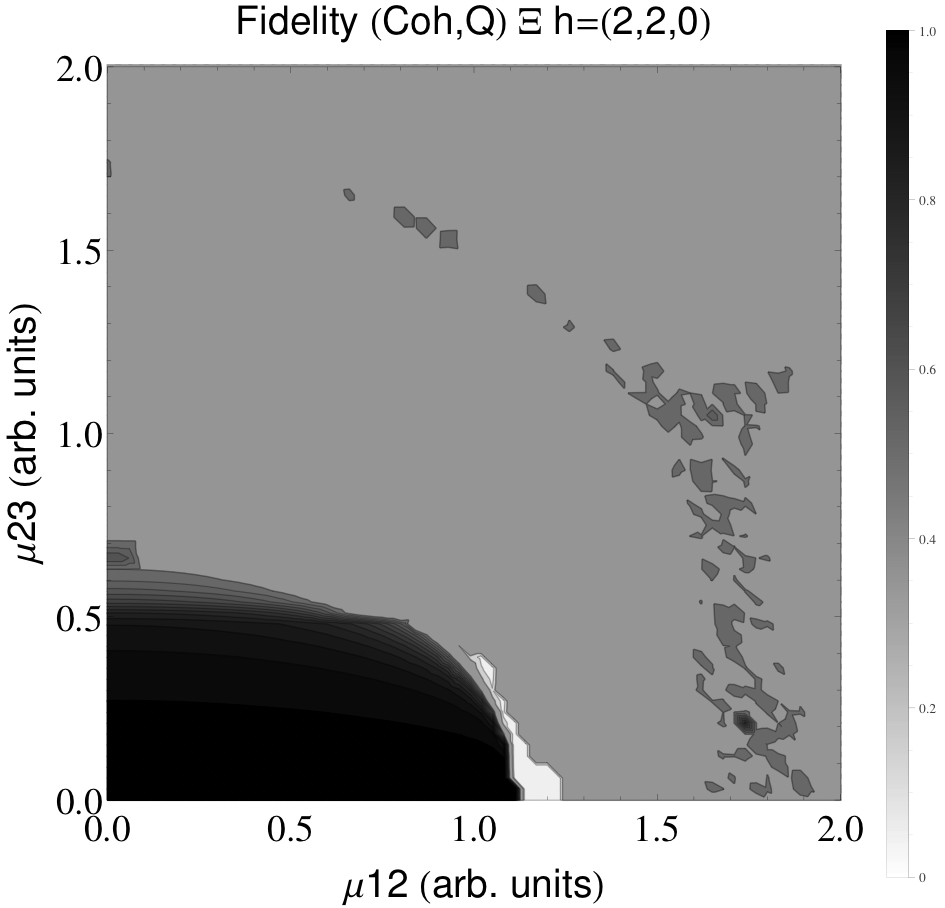}\quad{}\quad{}\quad{}\quad{}\includegraphics[scale=0.24]{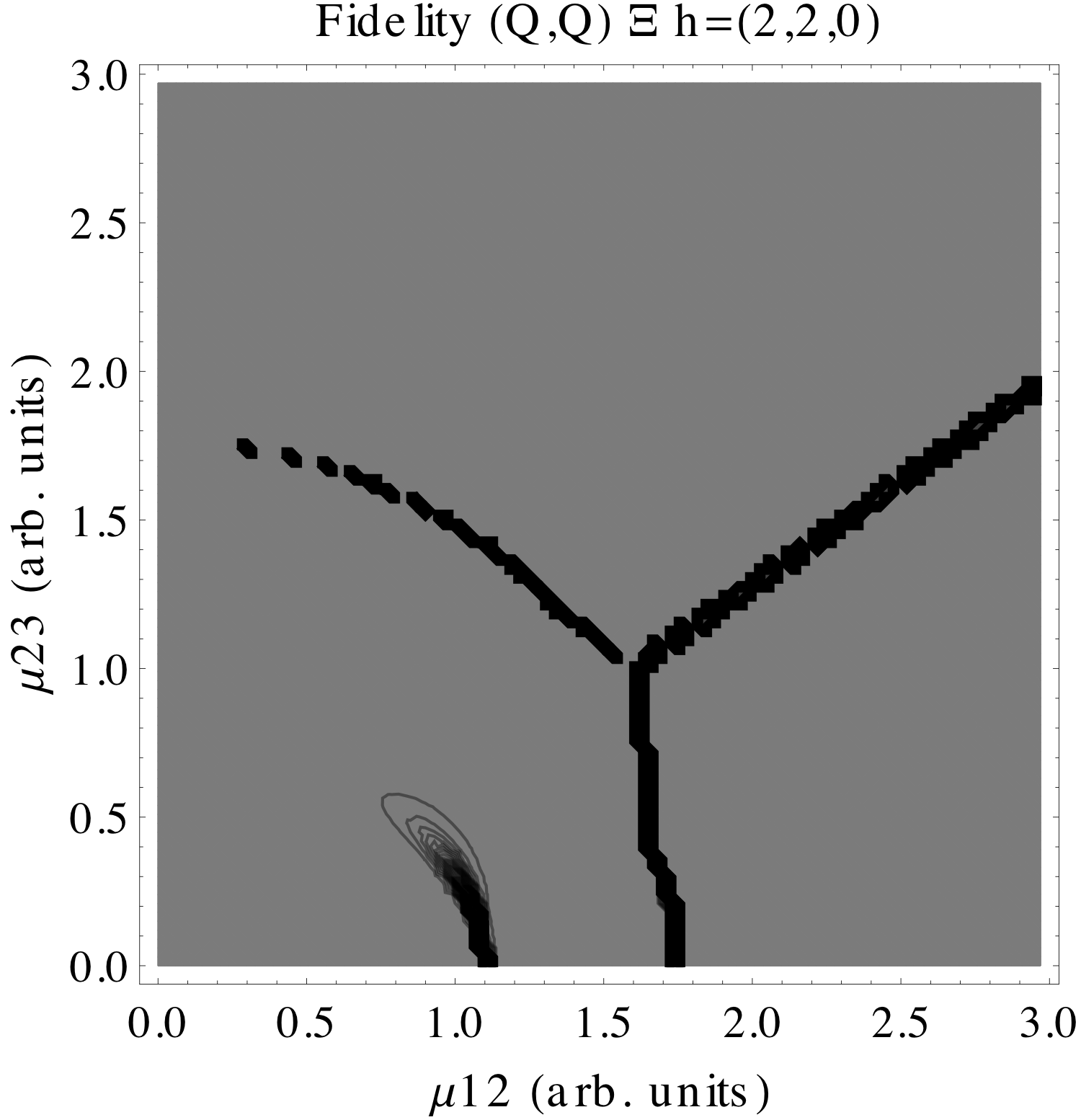}\quad{}
		\par\end{centering}
	
	\caption{Left: Contour plot of the fidelity between coherent states and quantum
		solution as a function of the coupling parameters $\mu_{12}$ and
		$\mu_{23}$, values range between 0 (white) and 1 (black). Right:
		Contour plot of the fidelity between neighboring quantum states as
		a function of the coupling parameters $\mu_{12}$ and $\mu_{23}$,
		black dots represent a drop in the fidelity below 1. Both figures
		were obtained using $\omega_{1}=1.\bar{3}$, $\omega_{1}=1.\bar{6}$,
		$\Omega=0.5$ and correspond to the $\Xi$ configuration in the $h=(2,2,0)$
		representation. Units are arbitrary but the same for all non-dimensionless
		quantities ($\hbar=1$). (Noise in the plots is due to numerical minimization; see text.)\label{fig:12}}
\end{figure*}

\begin{figure*}[t]
	\begin{centering}
		\includegraphics[scale=0.39]{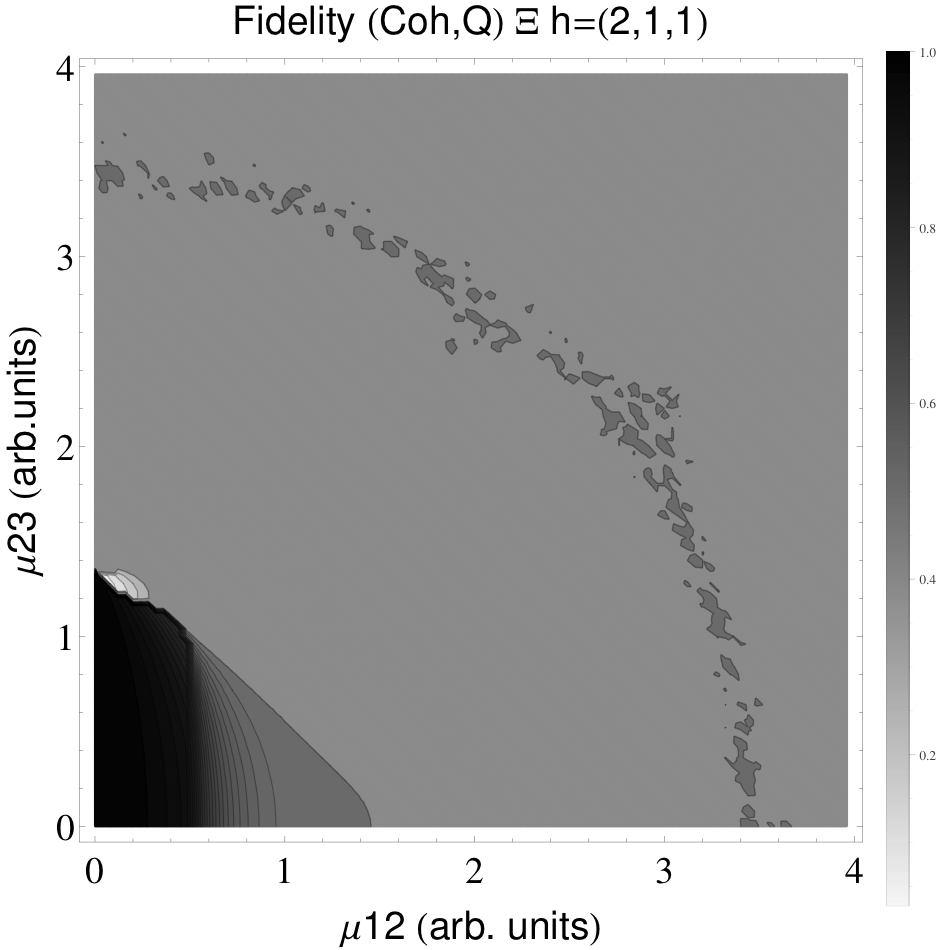}\quad{}\quad{}\quad{}\quad{}\includegraphics[scale=0.23]{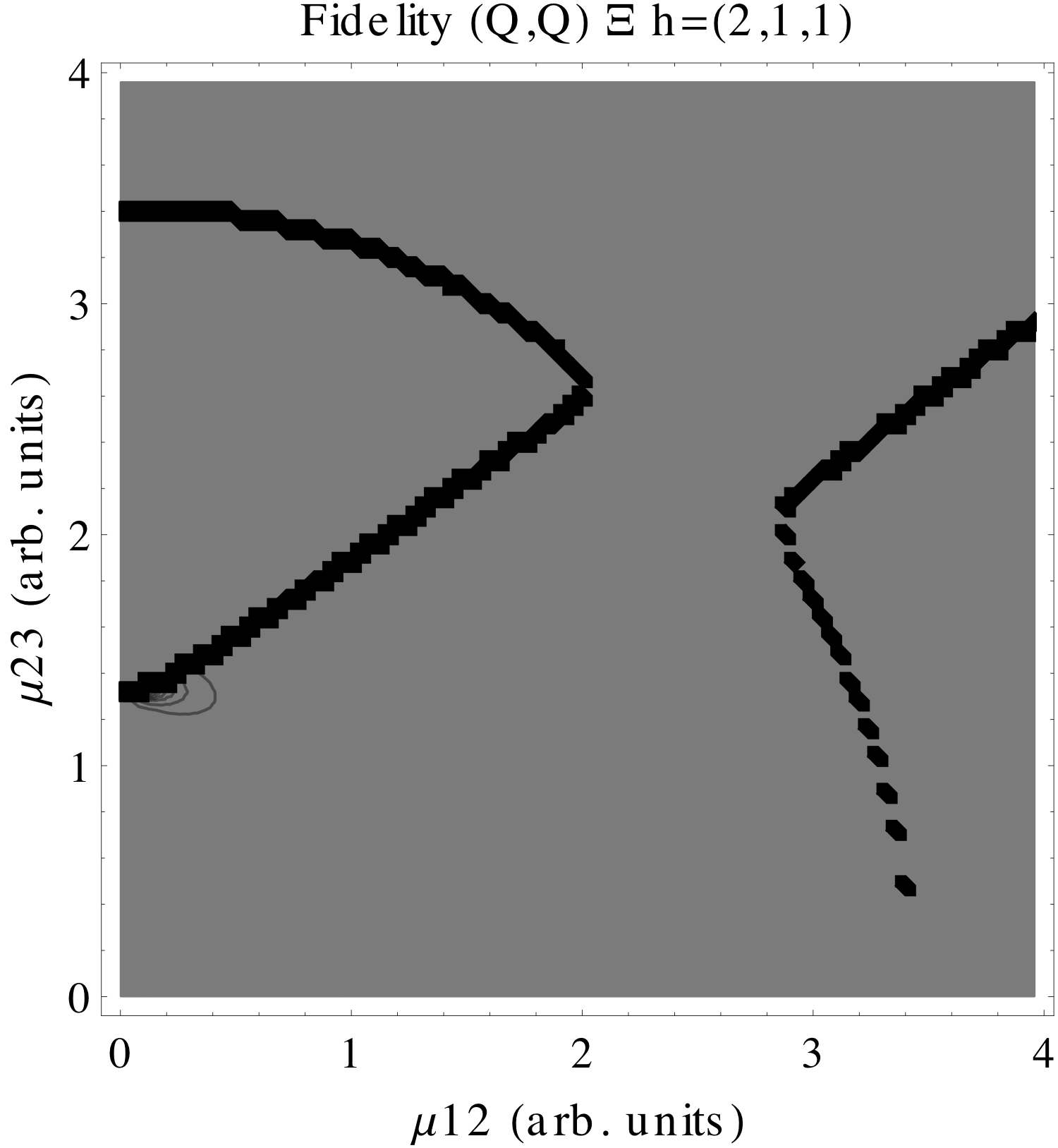}\quad{}
		\par\end{centering}
	
	\caption{Left: Contour plot of the fidelity between coherent states and quantum
		solution as a function of the coupling parameters $\mu_{12}$ and
		$\mu_{23}$, values range between 0 (white) and 1 (black). Right:
		Contour plot of the fidelity between neighboring quantum states as
		a function of the coupling parameters $\mu_{12}$ and $\mu_{23}$,
		black dots represent a drop in the fidelity below 1. Both figures
		were obtained using $\omega_{1}=1.\bar{3}$, $\omega_{1}=1.\bar{6}$,
		$\Omega=0.5$ and correspond to the $\Xi$ configuration in the $h=(2,1,1)$
		representation. Units are arbitrary but the same for all non-dimensionless
		quantities ($\hbar=1$). (Noise in the plots is due to numerical minimization; see text.)\label{fig:13}}
\end{figure*}


\subsubsection*{Fidelity between neighboring states, and quantum phase transitions}

Fidelity is a measure of the ``distance'' between two quantum states; given $\left|\phi\right\rangle $ and $\left|\varphi\right\rangle $
it is defined as

\begin{equation}
F(\phi,\varphi):=\left|\left\langle \phi|\varphi\right\rangle \right|^{2}.
\end{equation}

Across a QPT the ground state of a system changes drastically, thus it is natural to expect a drop in the fidelity between
neighboring states near the transition. This drop has been, in fact, already shown to happen for two and three-level systems \cite{key-28,key-29,key-30}.

Due to the above, the definition of the concept of quantum phase that we will be using throughout this paper is that of an open region in the space of parameters where the fidelity between neighboring states is close to $1$, therefore the QPT's are characterized by values of this fidelity close to $0$.

\section*{Methodology}

To study the QPT's in our system we need to know its ground state; in this work we use a variational approach and apply the energy surface minimization method to estimate it. This method consists on minimizing the surface that is obtained by taking the expectation value of the modeling hamiltonian with respect to some trial variational state. The strength of this method lies on the choice of the trial state, as it is the latter, after minimization, the one that will be modeling the ground state of the system.

Here we take a variational approach for both the matter and the radiation field, using a tensor product of $HW(1)$ coherent states (the usual coherent states of the harmonic oscillator) for the radiation field, and $SU(3)$ coherent states for the atomic field. As our system is not integrable, and the expression for the energy surface is unwieldy, the minimization is carried out numerically.

\subsubsection*{Coherent states of $HW(1)$}

For the electromagnetic field, the annihilation and creation operators $a$ and $a^{\dagger}$, appearing in the modeling hamiltonian (\ref{eq:2}), satisfy the commutation relations of the Lie algebra generators of the Heisenberg-Weyl group $HW(1)$:

\begin{center}
$\left[a,a^{\dagger}\right]=1$,
\end{center}
hence, a natural choice of trial states for the radiation field are the coherent states of $HW(1)$, defined as the application of the displacement operator to the radiation's lowest energy state: 

\begin{equation}
\left|\alpha\right\rangle :=e^{\alpha a^{\dagger}-\alpha^{*}a}\left|0\right\rangle =e^{-\frac{\left|\alpha\right|^{2}}{2}}\sum_{\nu=0}^{\infty}\frac{\alpha^{\nu}}{\sqrt{\nu!}}\left|\nu\right\rangle,
\label{eq:5}
\end{equation}
where $\left|\nu\right\rangle$ are the Fock states of the electromagnetic field.

\subsubsection*{Coherent states of $SU(3)$}

For the atomic field, as we have already mentioned, the operators $J_{z}^{\left(1\right)}$, $J_{z}^{\left(2\right)}$, $e_{12}$, $e_{23}$, $e_{12}^{\dagger}$ and $e_{23}^{\dagger}$ form a basis for the Lie algebra of $SU(3)$, thus, analogously as for the radiation field, it is natural to use the coherent states of $SU(3)$ as trial states; these are defined as the application of the exponential of the raising operators $e_{12}^{\dagger}$, $e_{23}^{\dagger}$ and $e_{13}^{\dagger}=\left[e_{23}^{\dagger},e_{12}^{\dagger}\right]$ to the atomic's lowest energy state, and in the Gelfand-Tsetlin scheme take the form:

\begin{center}
$\left|\gamma,h\right\} :=e^{\gamma_{3}e_{12}^{\dagger}+\gamma_{2}e_{13}^{\dagger}+\gamma_{1}e_{23}^{\dagger}}
\left|\begin{array}{ccc}
h_{1} & h_{2} & h_{3}\\
h_{1} & h_{2}\\
h_{1}
\end{array}\right\rangle$,
\end{center}
where the delimiters $\left|\cdot\right\}$ mean the state is not normalized. Performing this calculation gives us the following expression for the coherent states of $SU(3)$:

\begin{multline}
\left|\gamma,h\right\}={\displaystyle \sum_{n=0}^{h_{2}-h_{3}}}{\displaystyle\;\;\; \sum_{\ell=0}^{h_{1}-h_{2}}}{\displaystyle\;\;\; \sum_{m=0}^{h_{2}-h_{3}-n}}{\displaystyle\;\;\; \sum_{j=0}^{h_{1}-h_{2}-\ell+n}}\gamma_{1}^{n}\gamma_{2}^{\ell+m}\gamma_{3}^{j}\\\cdot\left(\begin{array}{c}
h_{2}-h_{3}\\
n
\end{array}\right)^{\frac{1}{2}}\left(\begin{array}{c}
h_{1}-h_{2}-\ell+n\\
j
\end{array}\right)^{\frac{1}{2}}\left(\begin{array}{c}
m+j\\
j
\end{array}\right)^{\frac{1}{2}}\\\cdot{\displaystyle \frac{S_{\ell mn}\left(h\right)}{\left(\ell+m\right)!}}\left|\begin{array}{ccc}
h_{1} & h_{2} & h_{3}\\
h_{1}-\ell & h_{2}-n-m\\
h_{1}-\ell-m-j
\end{array}\right\rangle.
\label{eq:6}
\end{multline}

Here, the numbers $S_{\ell mn}\left(h\right)$ are defined as the scalars obtained from the application of the operator $\left(e_{13}^{\dagger}\right)^{\ell+m}$ to the resulting states from the previous application of $e^{\gamma_{1}e_{23}^{\dagger}}$, namely:

\begin{multline}
\left(e_{13}^{\dagger}\right)^{\ell+m}\left|\begin{array}{ccc}
h_{1} & h_{2} & h_{3}\\
h_{1} & h_{2}-n\\
h_{1}
\end{array}\right\rangle = \\ S_{\ell mn}\left(h\right)\left|\begin{array}{ccc}
h_{1} & h_{2} & h_{3}\\
h_{1}-\ell & h_{2}-n-m\\
h_{1}-\ell-m
\end{array}\right\rangle
\end{multline}

\section*{Results and Discussion}

The results presented in the main body of this work correspond to the analysis made with the atoms of the system being in the $\Xi$ configuration. Results for the $\Lambda$ and $V$ configurations are shown in the supplemental material \cite{key-31}.

As it has already been stated, the energy surface minimization was carried out numerically; figures \ref{fig:2} and \ref{fig:3} show the results of this procedure. In them, the average ground-state's energy of the system is plotted as a function of the dipolar coupling parameters $\mu_{12}$ and $\mu_{23}$ for all the four possible representations and cooperation numbers available for $N=4$. It can be seen from these figures that the area of the normal region (shown in dark gray) in the $\mu_{ij}$ plane gets larger as $n_{c}$ gets smaller; this is consistent with the intuition behind the cooperation number as the fewer the effective number of atoms is, the stronger the required coupling needs to be for the system to reach the super-radiant phase.

Figures \ref{fig:4} and \ref{fig:5} display the average number of photons in the ground state of the system, which in the normal region is zero but grows rapidly as we go deeper into the super-radiant phase. This growth has been shown to be of fourth order with respect to the dipolar coupling parameters for two-level systems \cite{key-24}.

The atomic observables are studied in figures \ref{fig:6} to \ref{fig:9}, they show both the average of half the population difference between the second and first levels, and the average of half the population difference between the third and second levels, which correspond respectively to the expectation value, in the ground state, of the $Jz_{1}$ and $Jz_{2}$ operators. These figures reflect one of the features that make representation theory and the Gelfand-Tsetlin labeling scheme useful tools to describe this kind of systems: notice that the parameters $h_{1}$, $h_{2}$ and $h_{3}$ represent, respectively, the atomic population of the first, second and third level, in the normal region of the system.

As the methodology used in this work provides an approximation to the ground state, a comparison between this and the real quantum solution, calculated by explicitly diagonalizing the hamiltonian matrix, is presented in figures \ref{fig:10} to \ref{fig:13} by means of the fidelity between them $F(Coh,Q)$, along with the real QPTs obtained using the fidelity between neighboring quantum states $F(Q,Q)$, which we are using to characterize the real QPT.

It is worth mentioning that, in this case, there are mainly two ways in which we can calculate $F(Q,Q)$, one is to compare states across a horizontal line in the ($\mu_{12},\mu_{23}$) plane and the other is to do it across a vertical line, the first method being particularly sensible to vertical QPTs and the second method to horizontal ones. In this work, as both approaches looked almost identical, we decided to only show the resulting plots of one of them. It is important to point out, however, that this decision made continuous lines to look somehow dashed in some parts of our Fidelity (Q,Q) plots.

Some interesting characteristics of the system arise from the results displayed on these figures, the most notorious one being the fact that in the normal region $F(Coh,Q)\approx1$, meaning both solutions are nearly identical there. However, near the coherent QPT the fidelity starts falling rapidly until it reaches $F(Coh,Q)\approx\frac{1}{2}$ in the super-radiant region; this specific value is not a coincidence, it emerges from a mix of parities the coherent states carry, derived from a symmetry in the total number of excitations of the system. States that respect this symmetry (symmetry-adapted states, or SAS) can be constructed, and have actually already been used to study two- and three-level systems \cite{key-24,key-32} (the latter only for the symmetric representation), as well as other kind of systems \cite{key-33}.

Another interesting aspect of the system present in figures \ref{fig:10} to \ref{fig:13} is the traces of the coherent QPT present in the Fidelity (Q,Q) plots. These are more noticeable as the cooperation number increases. Traces of the real QPT in the Fidelity (Coh,Q) plots are expected, as this fidelity is literally comparing both kinds of states; however, to see a drop in $F(Q,Q)$ where the coherent QPT occurs is quite a remarkable feature, as there is, in principle, no information about the coherent state approximation in the Fidelity (Q,Q) plots. We attribute this phenomenon to the following two facts: it has been shown \cite{key-24}, for two-level systems, that the quantum and the SAS solution coincide in the cooperation-number thermodynamic limit (i.e. $n_{c}\longrightarrow\infty$); and both coherent and SAS solutions can be made to have the same normal region (minimizing both with the same critical values). This leads us to conclude that, as $n_{c}\longrightarrow\infty$, the traces of the coherent QPT gradually become the real QPT.

Lastly, when the rotating wave approximation is considered, the system has been shown to have a triple point for the symmetric representation \cite{key-34}, which is fixed in parameter's space ($\mu_{12},\mu_{23}$), is independent on the number of atoms, and prevails in the thermodynamic limit. This triple point also appears in our Fidelity (Q,Q) plots but, from all other figures analyzed, it does not seem to be relevant in the coherent approximation when the full hamiltonian is considered.

\section*{Conclusions}

In this work we showed the usefulness of representation theory and the Gelfand-Tsetlin labeling scheme to study systems of matter interacting with radiation in the dipolar approximation, allowing us to easily define the cooperation number and immediately knowing the atomic population of each level in its normal phase.

We see from the studied observables (energy, photon number, half the atomic population between second and first levels, and between third and second levels), presented in figures \ref{fig:2} to \ref{fig:9}, that the given definition of the cooperation number (\ref{eq:3}) is consistent with the intuition of an effective number of atoms in the system, mainly by the fact that the area of the normal region (according to the coherent approximation) gets larger as the cooperation number decreases.

The reliability of the coherent approximation was analyzed using the fidelity between the coherent and quantum solutions, shown in figures \ref{fig:10} to \ref{fig:13}, along with the real QPT via a drop in the fidelity between quantum neighboring states (Fidelity (Q,Q) plots). In the latter case, traces of the coherent QPT were observed regardless of the fact that the fidelity F(Q,Q) was calculated using just the quantum solution, a characteristic we attributed, based on previous results obtained for two-level systems, to the fact that both solutions coincide in the cooperation-number thermodynamic limit.

In conclusion, we utilized a coherent approximation to the system's ground state to study its quantum phase transitions, which we used to justify the given definition of cooperation number, showing how this affects the behavior of the relevant observables of the system near the transitions for all configurations of three-level atoms.

\begin{acknowledgments}
This work was partially supported by DGAPA-UNAM under
project IN101217. L.F. Quezada thanks CONACyT-M\'exico for financial support
(Grant \#379975).
\end{acknowledgments}


\begin{thebibliography}{10}
\bibitem{key-1}
R.H. Dicke, \href{http://journals.aps.org/pr/abstract/10.1103/PhysRev.93.99}{Phys. Rev. 93, 99 (1954)}.
\bibitem{key-9}
H.I. Yoo and J.H. Eberly, \href{https://www.sciencedirect.com/science/article/pii/0370157385900158}{\em Phys. Rep. 118, 239 (1985)}.
\bibitem{key-10}
O. Civitarese and M. Reboiro, \href{https://www.sciencedirect.com/science/article/pii/S037596010600613X}{Phys. Lett. A 357, 224 (206)}.
\bibitem{key-11}
N.H. Abdel-Wahab, \href{http://iopscience.iop.org/article/10.1088/0031-8949/76/3/006}{Phys. Scr. 76, 244 (2007)}.
\bibitem{key-12}
N.H. Abdel-Wahab, \href{http://www.worldscientific.com/doi/abs/10.1142/S0217984908016868}{\em Mod. Phys. Lett. B 22, 2587 (2008)}.
\bibitem{key-13}
M Hayn, C. Emary and T. Brandes, \href{https://journals.aps.org/pra/abstract/10.1103/PhysRevA.84.053856}{Phys. Rev. A 84, 053856 (2011)}.
\bibitem{key-14}
M. Hayn, C. Emary and T. Brandes, \href{https://journals.aps.org/pra/abstract/10.1103/PhysRevA.86.063822}{Phys. Rev. A 86, 063822 (2012)}.
\bibitem{key-15}
S. Cordero, R. L\'opez-Pe\~na, O. Casta\~nos and E. Nahmad-Achar, \href{https://journals.aps.org/pra/abstract/10.1103/PhysRevA.87.023805}{Phys. Rev. A 87, 023805 (2013)}.
\bibitem{key-16}
S. Cordero, O. Casta\~nos, R. L\'opez-Pe\~na and E. Nahmad-Achar, E. \href{http://iopscience.iop.org/article/10.1088/1751-8113/46/50/505302}{J. Phys. A 46, 505302 (2013)}.
\bibitem{key-17}
S. Cordero, E. Nahmad-Achar, R. L\'opez-Pe\~na and O. Casta\~nos, \href{https://journals.aps.org/pra/abstract/10.1103/PhysRevA.92.053843}{Phys. Rev. A 92, 053843 (2015)}.
\bibitem{key-18}
S. Cordero, O. Casta\~nos, R. L\'opez-Pe\~na and E. Nahmad-Achar, \href{https://journals.aps.org/pra/abstract/10.1103/PhysRevA.94.013802}{Phys. Rev. A 94, 013802 (2016)}.
\bibitem{key-19}
V.N. Chernega, O.V. Manko and V.I. Manko, \href{https://link.springer.com/article/10.1007/s10946-017-9662-4}{J. Russ. Laser Res. 38, 416 (2017)}.
\bibitem{key-20}
A.E. Kozhekin, K. M{\o}lmer and E. Polzik, \href{https://journals.aps.org/pra/abstract/10.1103/PhysRevA.62.033809}{Phys. Rev. A 62, 033809 (2000)}.
\bibitem{key-21}
A. Gorshkov, A. Andr\'e, M. Fleischhauer, A. S{\o}rensen and M. Lukin, \href{https://journals.aps.org/prl/abstract/10.1103/PhysRevLett.98.123601}{Phys. Rev. Lett. 98, 123601 (2007)}.
\bibitem{key-22}
J. Nunn, I.A. Walmsley, M.G. Raymer, K. Surmacz, F.C. Waldermann, Z. Wang and D. Jaksch \href{https://journals.aps.org/pra/abstract/10.1103/PhysRevA.75.011401}{Phys. Rev. A 75, 011401 (2007)}.
\bibitem{key-23}
J.L. Morton, A.M. Tyryshkin, R.M. Brown, S. Shankar, B.W. Lovett, A. Ardavan, T. Schenkel, E.E. Haller, J.W. Ager, and S.A. Lyon, \href{http://www.nature.com/articles/nature07295}{\em Nature 455, 1085 (2008)}.
\bibitem{key-24}
L.F. Quezada and E. Nahmad-Achar, \href{https://journals.aps.org/pra/abstract/10.1103/PhysRevA.95.013849}{Phys. Rev. A 95, 013849 (2017)}.
\bibitem{key-25}
L.F. Quezada and E. Nahmad-Achar, \href{http://www.mdpi.com/1099-4300/20/2/72}{Entropy 20, 72 (2018)}.
\bibitem{key-26}
I.M. Gelfand and M.L. Tsetlin, \href{http://inspirehep.net/record/1402918}{Dokl. Akad. Nauk 71, 825 (1950)}.
\bibitem{key-27}
A. Arne, M. Kalus, A. Huckleberry and J.V. Delft, \href{http://aip.scitation.org/doi/10.1063/1.3521562}{J. Math. Phys. 52, 023507 (2011)}.
\bibitem{key-28}
E. Nahmad-Achar, S. Cordero, O. Casta\~nos and R. L\'opez-Pe\~na, \href{http://iopscience.iop.org/article/10.1088/0031-8949/90/7/074026}{Phys. Scr. 90, 074026 (2015)}.
\bibitem{key-29}
P. Zanardi and N. Paunkovi, \href{https://journals.aps.org/pre/abstract/10.1103/PhysRevE.74.031123}{Phys. Rev. E 74, 031123 (2006)}.
\bibitem{key-30}
O. Casta\~nos, E. Nahmad-Achar, R. L\'opez-Pe\~na and J.G. Hirsch, \href{https://journals.aps.org/pra/abstract/10.1103/PhysRevA.86.023814}{Phys. Rev. A 86, 023814 (2012)}.
\bibitem{key-31}
Supplemental Material ({\color{red}{Editor: please insert here appropriate link for the Supplemental Material}}).
\bibitem{key-32}
R. L\'opez-Pe\~na, S. Cordero, E. Nahmad-Achar and O. Casta\~nos, \href{http://iopscience.iop.org/article/10.1088/0031-8949/90/6/068016}{Phys. Scr. 90, 068016 (2015)}.
\bibitem{key-33}
M. Calixto, E. Romera and R. del Real, \href{http://iopscience.iop.org/article/10.1088/1751-8113/45/36/365301}{J. Phys. A 45, 365301(2012)}.
\bibitem{key-34}
E. Nahmad-Achar, S. Cordero, R. L\'opez-Pe\~na and O. Casta\~nos, \href{http://iopscience.iop.org/article/10.1088/1751-8113/47/45/455301}{J. Phys. A 47, 455301 (2014)}.
\end{thebibliography}
\end{document}